\documentclass[11pt,a4paper]{article}
\usepackage[utf8]{inputenc}
\usepackage[left=2.5cm,top=3cm,right=2cm,bottom=3cm]{geometry}
\usepackage{authblk}
\usepackage{etex}
\usepackage[T1]{fontenc}
\usepackage{amssymb,amsmath,geometry}
\usepackage{mathtools}
\usepackage{graphicx}
\usepackage{fancyhdr}
\usepackage{fancybox}
\usepackage{shadow}
\usepackage{empheq}
\usepackage{titlesec}
\usepackage{titletoc}
\usepackage{soul, color}
\usepackage{float}
\usepackage{shorttoc}
\usepackage{xcolor}
\usepackage{fancybox}
\usepackage{array}
\usepackage{tabularx}
\usepackage{booktabs}
\usepackage{rotating}
\usepackage{bigstrut}
\usepackage{aeguill}

\usepackage{changepage}
\newenvironment{changemargin}[2]{\begin{list}{}{%
\setlength{\topsep}{0pt}%
\setlength{\leftmargin}{0pt}%
\setlength{\rightmargin}{0pt}%
\setlength{\listparindent}{\parindent}%
\setlength{\itemindent}{\parindent}%
\setlength{\parsep}{0pt plus 1pt}%
\addtolength{\leftmargin}{#1}%
\addtolength{\rightmargin}{#2}%
}\item }{\end{list}}

\usepackage{hyperref}
\def\sym#1{\ifmmode^{#1}\else\(^{#1}\)\fi}
\usepackage{caption}
\usepackage{blindtext}
\usepackage[utf8]{inputenc}


\usepackage{caption}
\usepackage{subcaption}
\usepackage{mathtools}

\usepackage{graphicx}

\usepackage{enumerate}

\usepackage{hyperref}
\hypersetup{backref, colorlinks=true, citecolor=blue!50!black}
\usepackage{graphicx}

\usepackage{lscape} 
\renewcommand{\thetable}{\Roman{table}}
\renewcommand{\thefigure}{\Roman{figure}}

\usepackage{titlesec}
\usepackage{setspace} 
\usepackage{indentfirst}

\makeatletter
  \newcommand\smalls{\@setfontsize\smalls{10.3pt}{6}}
\makeatother

\makeatletter
  \newcommand\footnotesizes{\@setfontsize\footnotesizes{9.6pt}{6}}
\makeatother

\usepackage[authoryear]{natbib}
\newsavebox\tmpbox

\usepackage{caption}
\usepackage{subcaption}

\begin{document}

\title{Institutions, Education, and Religious Change: \\ Evidence from Colombia}

\author[1]{Hector Galindo-Silva\thanks{galindoh@javeriana.edu.co}}
\author[1]{Paula Herrera-Id\'arraga\thanks{pherrera@javeriana.edu.co}}
\affil[1]{\small Department of Economics, Pontificia Universidad Javeriana}

\maketitle
\vspace{-2pt}

\begin{abstract}
How do religious identities change? We study the effects of civic education reforms on religious identification using Colombia’s 1991 Constitution, which dismantled the country’s confessional regime and mandated constitutional instruction in high schools. Exploiting cohort-based variation in exposure to the reform and nationally representative survey data, we implement a difference-in-differences design. We find that exposure to the constitutional curriculum reduced Catholic self-identification by about three percentage points. This decline reflects a reallocation of religious identities rather than a generalized decline in religiosity. In regions where Catholic institutional presence was historically weaker, Catholic losses translate into switching toward non-Catholic Christian denominations and higher religious attendance. In contrast, in regions where Catholic dominance was stronger, the decline is associated with increased secular identification and lower attendance. These patterns hold across ethnic and non-ethnic groups and are shaped primarily by regional religious supply rather than ethnicity per se. Overall, the results show that civic education can reconfigure religious identities by reshaping the relative legitimacy of competing affiliations.
\end{abstract}


\newpage

\section{Introduction}

The global religious landscape has shifted markedly in recent decades. Across countries, traditional forms of religiosity have declined in many high-income societies, while remaining stable or even expanding in several low-income contexts \citep{inglehart2020religion}. At the same time, growing evidence suggests that these changes do not imply a uniform process of secularization. Instead, religious transformation often takes the form of diversification, adaptation, and institutional reorganization, with new denominations and identities emerging alongside—rather than replacing—established traditions \citep{lowes2025religion}.

These transformations have important implications, as religious identity shapes individual behavior and a wide range of social outcomes, including savings, labor supply, education, gender norms, political attitudes, and household decision-making \citep{becker2024religion, guiso2003people, benjamin2016religious, valencia2019mission, kersting2020weber, voigt2024determinants}. Accordingly, identifying the forces that reshape religious identity is essential for understanding broader patterns of social change and inclusive development.

Latin America provides a salient setting in which these global patterns of religious transformation are especially pronounced. Catholicism historically structured public life in the region, shaping political institutions, education systems, and social norms. In Colombia, the 1886 Constitution formalized this dominance by establishing Catholicism as the state religion and granting it privileged authority over schooling, family law, and moral regulation. This confessional order closely intertwined Catholic and national identities, while other denominations remained marginal. By the late twentieth century, however, Colombia—like much of Latin America—experienced sustained declines in Catholic affiliation alongside the expansion of Evangelical, Pentecostal, and secular identities, reflecting a broader process of religious reconfiguration driven by demographic change, organizational expansion, and institutional reform.

We focus on one such institutional reform: Colombia’s 1991 Constitution. The new charter dismantled the country’s confessional regime by guaranteeing freedom of religion and belief and redefining the relationship between religion and the state. Crucially, it also introduced a civic education mandate requiring secondary schools to teach the Constitution’s principles. Beginning in 1994, these courses became a graduation requirement, systematically exposing cohorts of students to norms of equality, tolerance, and religious pluralism at a formative stage of identity formation.

We examine whether this reform altered religious identity. Specifically, we test whether exposure to mandatory courses on the 1991 Constitution reduced the likelihood that individuals self-identify as Catholic. Our central hypothesis is that, by weakening Catholicism’s symbolic status as the country’s default religion and legitimizing religious pluralism, the reform lowered the social and normative costs of alternative affiliations, leading to a reconfiguration of religious identities.

We identify the effect using a difference-in-differences design that exploits the staggered introduction of the constitutional curriculum in 1994. Treatment is defined by completing high school in 1994 or later, when the new courses became mandatory, while individuals who completed schooling or dropped out prior to 1994 constitute the control group. Comparing these cohorts in nationally representative survey data allows us to estimate the causal impact of exposure to the reform on religious identification.

Our baseline results show that exposure to constitutional instruction reduced the probability of identifying as Catholic by about 2–3 percentage points, a meaningful shift in a context where Catholicism is the dominant affiliation. The estimates are statistically significant across specifications, robust to alternative age windows, and consistent with the identifying assumption of parallel trends: placebo tests using pre-1994 cohorts reveal no systematic pre-existing differences between treated and untreated groups.

Additional evidence sheds light on the mechanisms behind this change. The decline in Catholic identification masks distinct regional reallocations. In the Pacific and Atlantic regions—where Catholic institutional presence has historically been weaker—Catholic losses are accompanied by increases in identification with non-Catholic Christian groups, with little change in secular affiliation. In contrast, in the Andean core, where Catholic dominance was stronger, the decline in Catholicism is associated with a rise in secular identification but not with shifts toward other Christian denominations. Religious practice adjusts accordingly: attendance increases in peripheral regions and declines in the core, consistent with substitution toward more practice-intensive denominations in the former and weakened overall religiosity in the latter. Using self-reported ethnic identity, we further show that these patterns are mediated, but not driven, by ethnic segregation. Catholic identification declines across all ethnic groups, while denominational switching and secularization follow the regional institutional structure rather than ethnic lines. Taken together, these results support the interpretation that civic education on the 1991 Constitution eroded Catholicism’s symbolic monopoly and legitimized alternative identities, with the direction of switching shaped by pre-existing religious supply.

We also consider alternative explanations. There is no evidence that the reform reduced religiosity overall: aggregate religious attendance does not decline, and among ethnic minorities it increases, indicating redirection rather than withdrawal. Effects on political ideology are small and statistically insignificant, and labor-force participation is unchanged. Overall, the evidence reinforces our preferred interpretation: the curriculum legitimized denominational pluralism and made both switching toward non-Catholic denominations and secular identification more socially acceptable, without altering ideology, economic status, or general religious engagement.

This paper contributes to a growing literature that studies culture as a system of values and identities shaping economic and political behavior through long-run persistence and intergenerational transmission \citep{GuisoSapienzaZingales2006, GuisoSapienzaZingales2009, FernándezFogli2009}. While much of this work emphasizes deep historical roots, recent research shows that cultural traits can respond to institutional change: political reforms, legal transformations, and major social shocks can alter norms and identities, sometimes rapidly \citep{AlganCahuc2010, fernandez2013cultural, Bau2021}. Education has emerged as a key channel through which constitutions transmit new civic norms \citep{galindo2025constitutions}. We extend this line of research by showing that religious identity—often viewed as a particularly persistent cultural marker—can also respond to institutional and educational change.

Within this framework, the paper makes three main contributions. First, it provides causal evidence that a contemporary constitutional reform—transmitted through mandatory civic education—can reshape religious identities, complementing a large literature that emphasizes the historical persistence of culture and religion in economic outcomes \citep{GuisoSapienzaZingales2006, Nunn2012, BeckerWoessmann2009}. Second, it contributes to the economics of religion and religious competition by identifying an institutional mechanism through which pluralism operates: the removal of Catholic constitutional privilege and the embedding of religious freedom into schooling lowered the perceived costs of switching religious affiliation. This mechanism is consistent with models of religions as competing platforms \citep{seabright2024divine} and with empirical evidence showing that education, state neutrality, and pluralistic environments facilitate religious conversion and denominational change \citep{barro2024religious, cantoni2015adopting, gruber2005religious}. Third, the regional heterogeneity we document highlights that the effects of civic and legal reforms depend critically on pre-existing religious supply and institutional density, echoing a broader insight in political economy that institutional shocks interact with local historical and organizational conditions \citep{AcemogluJohnsonRobinson2001, NunnWantchekon2011}. Together, these results underscore the role of education as a transmission channel through which institutional change shapes durable, but context-dependent, cultural outcomes.

The remainder of the paper is organized as follows. Section~\ref{sec_background} provides institutional and historical background relevant to the analysis. Section~\ref{sec_data} describes the data and empirical strategy. Section~\ref{sec_results} presents the main results. Section~\ref{sec_mechanisms} discusses the underlying mechanisms, while Section~\ref{section_robustness} reports robustness checks. Section~\ref{sec_conclusion} concludes.

\section{Background}\label{sec_background}

For more than a century prior to 1991, Colombia was formally a confessional state under the Constitution of 1886, which established Catholicism as the official religion. The Catholic Church exerted substantial influence not only over religious life but also over education, family law, and politics. This close Church–State alliance fostered a setting in which Catholic identity was widely perceived as synonymous with national identity, a configuration often described as \emph{nacional–catolicismo} in the Colombian context \citep{Arias2003, Beltran2013}. Although alternative faiths were legally permitted following the 1887 Concordat with the Vatican, they remained socially marginalized and enjoyed limited institutional recognition. Protestant denominations and other minority faith communities, in particular, faced persistent social stigma and weak organizational support \citep{Arias2003, Beltran2013}.

By the late twentieth century, this confessional framework came under increasing strain. Prolonged internal armed conflict, together with growing demands for expanded civil liberties, political inclusion, and institutional reform, generated pressure for a comprehensive constitutional overhaul. In response, the 1991 National Constituent Assembly was convened with the explicit aim of establishing a new social pact that would broaden democratic participation, address longstanding sources of conflict, and strengthen protections for historically marginalized groups. A central departure from the 1886 Constitution was the explicit constitutional recognition of religious diversity and state neutrality in matters of belief.

The 1991 Constitution dismantled Colombia’s confessional regime and replaced it with a secular framework grounded in individual rights. Article 18 guaranteed freedom of thought, conscience, and belief, while Article 19 established freedom of religion and equal legal protection for all faiths, declaring that all individuals may freely profess and disseminate their religion, individually or collectively, and that all churches are equal before the law. These provisions marked a decisive break from the prior constitutional order, under which Catholicism enjoyed privileged legal status and de facto regulatory authority over large domains of social life.

Beyond these legal guarantees, the Constitution also mandated civic education as a core component of schooling. Article 41 required that the study of the Constitution and civic education be compulsory in all public and private educational institutions. This mandate was operationalized through Law 107 of 1994, which required fifty hours of instruction on the 1991 Constitution as a prerequisite for high school graduation, and reinforced by Law 115 of 1994 (the General Education Law), which integrated civic education throughout the curriculum. Official pedagogical materials produced by the Ministry of Education emphasized principles of religious freedom, tolerance, and pluralism, explicitly signaling that Catholicism was no longer constitutionally privileged but one among many legitimate religious options.

These reforms constituted both a legal and educational shock to long-standing religious arrangements. For generations, Colombian students had been socialized under the assumption that Catholic identity was natural and dominant, reinforced by Church influence over schooling and moral instruction. In contrast, the post-1991 civic curriculum conveyed the message that religious affiliation was a matter of individual choice and that pluralism itself was a constitutional value. By redefining the boundaries of legitimate religious expression and embedding these principles in formal schooling, the reform plausibly weakened Catholicism’s symbolic monopoly and expanded the set of socially acceptable religious and non-religious identities, including Protestant denominations, Evangelical churches, and secular affiliations.

The broader historical context underscores the novelty of this reform. While several Latin American countries had already moved toward secular constitutions or curtailed Church privileges earlier in the twentieth century, Colombia’s transition occurred relatively late and followed a long period of close Church–State alignment. Importantly, the Colombian case combined formal constitutional guarantees with systematic diffusion through the education system, targeting adolescents at a formative stage of identity development. In this sense, the 1991 Constitution not only altered the legal status of religion but also actively transmitted new civic norms through schooling.

\medskip

These institutional changes interacted with a highly uneven religious landscape. Prior to 1991, Catholic institutional presence varied sharply across regions. The Church’s organizational footprint was strongest in Bogota and the Central and Eastern regions, where it was closely intertwined with state-building, public education, and the traditional party system under the 1886 Constitution and the 1887 Concordat \citep{Beltran2013, Arias2003}. By contrast, the Caribbean and Pacific regions historically exhibited weaker Catholic dominance and greater ethnic and cultural diversity. In the Caribbean, early Protestant influences are well documented, particularly in San Andrés and Providencia, where the Baptist Church was established as early as 1845 and consolidated as the majority denomination \citep{Beltran2013}. More broadly, Protestant and Evangelical missions frequently targeted peripheral populations.\footnote{For example, the \emph{Unión Misionera Evangélica} arrived in 1908 with an explicit focus on rural, Afro-descendant, and Indigenous communities, followed by additional missionary organizations in subsequent decades \citep{Beltran2013}.} These groups often concentrated in coastal cities such as Cartagena and Santa Marta–Ciénaga, as well as in Pacific and southwestern cities like Cali and Pasto, reinforcing patterns of non-Catholic growth in regions characterized by weaker state and Church presence.\footnote{Complementing this historical evidence, contemporary scholarship emphasizes that in peripheral regions—often marked by displacement, informality, and limited state capacity—conversion to new religious movements, especially Pentecostalism, has functioned as a response to social exclusion by providing intermediate institutions, social support, and moral order \citep{Beltran2013TX}. The Atlantic coast also includes a distinct Muslim center in Maicao (La Guajira), rooted in mid-twentieth-century Arab migration \citep{Beltran2013TX}. In these contexts, religious change has frequently taken the form of denominational reconfiguration rather than outright secularization.}

In sum, the 1991 Colombian Constitution represents a watershed in the country’s religious history. It redefined Colombia as a secular and pluralist state, constitutionally guaranteed freedom of religion, and embedded these principles within the education system through mandatory civic instruction. At the same time, Colombia’s religious field remained deeply heterogeneous across regions. Together, these features provide the institutional and historical context for our analysis of whether exposure to post-1991 civic education is associated with changes in religious self-identification.

\section{Data and Empirical Strategy}
\label{sec_data}

\subsection{Data}

Our main source of data is the \textit{Political Culture Survey} (\textit{Encuesta de Cultura Política}), conducted biennially by Colombia’s National Administrative Department of Statistics (DANE). This nationally representative survey covers adults aged 18 and older, stratified by region.\footnote{The regions are 1.Bogota, 2.Caribe, 3.Oriental, 4.Central, 5.Pacífica} Each wave includes approximately 36,000 respondents and collects information on political, social, and cultural attitudes, including trust in institutions, civic participation, and identity.  

In the 2015, 2017, 2019, and 2021 waves, the \textit{Political Culture Survey} included a question central to this study, asking respondents about their religious affiliation, which-as we discuss in the next section-constitutes our main outcome.\footnote{The exact survey question is: ``Si usted es de alguna religión, ¿puede decirme cuál es su religión?'' (If you belong to a religion, could you tell me which one?).} Respondents could choose among ten categories, and the vast majority answered. The options are: (i) Catholic, (ii) Protestant/Traditional Protestant, (iii) Evangelical or Pentecostal, (iv) Jehovah’s Witness, (v) Mormon (Church of Jesus Christ of Latter-Day Saints), (vi) Eastern non-Christian religions, (vii) Judaism, (viii) traditional religions, (ix) none, and (x) agnostic/atheist.

\begin{figure}[H]
\begin{center}
\caption{Evolution of Religious Adherence in Colombia, 1996-2023}
\label{fig_evolreladh}
\vspace{-0.3cm}
\includegraphics[width=14cm,height=10cm]{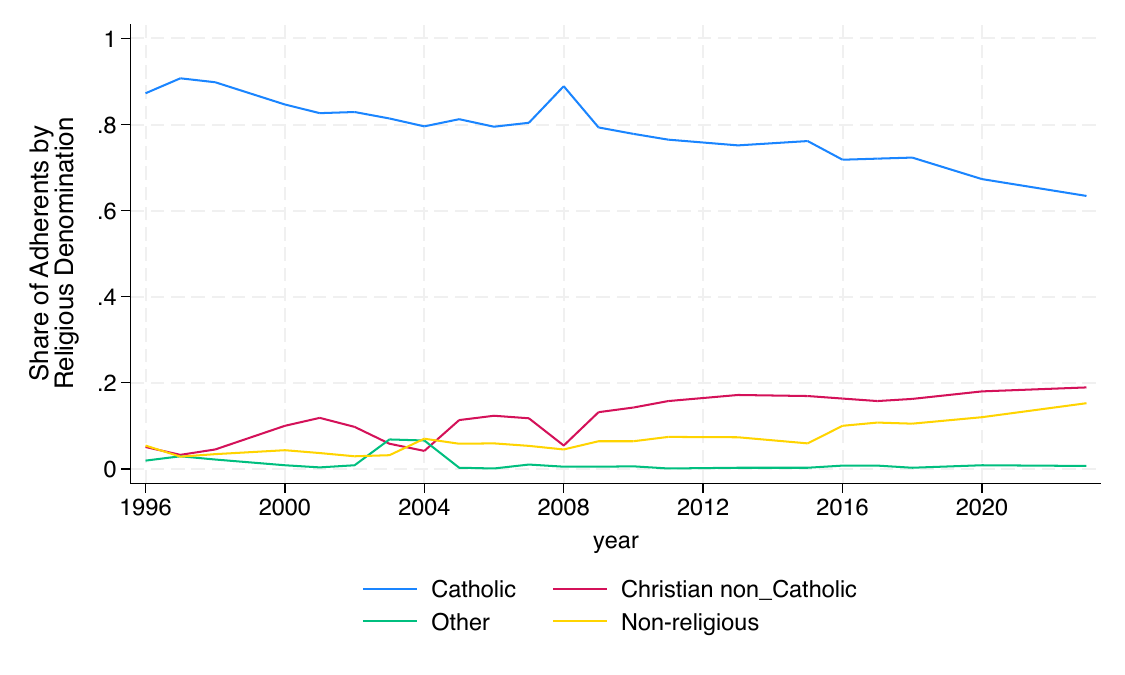}
 \begin{minipage}{14cm} \scriptsize\textbf{Note}: This figure uses data from the Latinobar\'ometro for Colombia, covering all available waves between 1996 and 2023. It plots the proportion of responses to the question, ‘What is your religion?’ where “Christian non-Catholic” includes the following denominations: Evangelical, Baptist, Methodist, Pentecostal, Adventist, Jehovah’s Witness, and Protestant. ‘Non- religious’ includes responses such as Agnostic and Atheist.
\end{minipage}
\end{center}
\end{figure}

Table~\ref{table_distributionreligident} reports the distribution across categories, showing that about 81\% of respondents identify as Catholic, while the remainder divide between non-Catholic Christians (about 11\%), those with no religion or atheist (about 6\%), and adherents of non-Christian religions (about 2\%). These percentages closely match the long-run trends presented in Figure~\ref{fig_evolreladh} for 1996--2023, which are constructed from a smaller but independent dataset available for a longer time span.   

The survey also collects demographic and socioeconomic characteristics (age, gender, ethnicity, education, and region of residence). We use these both to construct our econometric model---defining treatment and control groups based on respondents’ age and high school completion status---and as covariates to explore heterogeneity in treatment effects.  

\medskip

Despite its scale and comprehensiveness, the \textit{Political Culture Survey} presents two main limitations. First, it does not record the month of interview, which prevents us from precisely identifying respondents’ age at the time of the survey (i.e., in years and months). This is relevant because our empirical strategy requires as accurate an age measure as possible. As explained in the next section, we rely only on year of birth, a conservative approach that may bias our estimates toward zero. Second, the survey is only available from 2015 onward, whereas the treatment took place in 1994. This means outcomes are observed more than two decades after the reform. While this gap opens the possibility of confounding factors, it also highlights persistence: if we detect effects after such a long period, this suggests the curricular reform had lasting consequences.  

\medskip

In addition to the \textit{Political Culture Survey}, we use several complementary data sources to characterize the pre-reform religious and demographic environment. Information on Catholic institutional presence comes from municipal-level counts of Catholic churches in 1995, compiled by \textit{Fundación Social}. Data on ethnic composition are drawn from the 1993 Population Census conducted by DANE, which identifies Afro-descendant and Indigenous populations at the municipal level. We also use administrative data on population size, municipal finances, and living conditions as additional controls in descriptive analyses. These auxiliary datasets are used to document spatial heterogeneity and to explore potential mechanisms.

\subsection{Empirical Strategy}

Our empirical strategy builds on the staggered exposure to the 1991 constitutional curriculum across birth cohorts, as outlined in the introduction. Following the 1991 constitutional reform, Article 41 mandated civic education on the Constitution in all schools. Law 107 (1994) required completion of at least 50 hours of constitutional studies as a prerequisite for high school graduation, while Law 115 (1994) reinforced the inclusion of civic education across the curriculum. Thus, beginning in 1994, students in their final year of high school (grade 11) were systematically exposed to the new curriculum.\footnote{As noted in Section \ref{sec_background}, while the curriculum was formally mandated in 1994, its implementation was gradual. Anecdotal evidence suggests that coverage expanded over time through improved textbooks, pedagogical innovations, and stronger institutional support. Later cohorts may therefore have received more intensive exposure. Nevertheless, all individuals classified as treated (i.e., those aged 16 or younger in 1994) would have received at least some exposure to the reform. Those who completed high school likely encountered more sustained and comprehensive instruction. In next section, we examine heterogeneity across later cohorts.} Our approach is to compare outcomes between individuals plausibly exposed to these courses and those who were not.

We focus on individuals who were aged 16 or younger in 1994, the year the constitutional curriculum was introduced. We select this threshold because, in Colombia, students typically begin their last year of secondary school at age 16.\footnote{The Colombian education system consists of preschool, basic education (primary grades 1–5 and lower secondary grades 6–9), and upper secondary (grades 10–11), culminating in the high school diploma (\emph{bachiller}). Assuming students start school at age 5 and progress without grade repetition, they typically enter grade 11 at age 16 and graduate around age 17. Ministry of Education statistics from 2018 onward confirm that the average graduation age is close to 17, with a difference of approximately six months between public and private schools. We find no reason to believe this gap would have been significantly larger in earlier years. This supports our use of age 16 in 1994 as the treatment cutoff.} In our survey data, this corresponds to individuals aged 39 or younger in 2017, 41 or younger in 2019, and 43 or younger in 2021. For those in this group who completed high school, at least one year-long course on the 1991 Constitution would have been mandatory. We refer to this group as the affected cohort. In contrast, individuals older than 16 in 1994 were not exposed to the curriculum and constitute the unaffected cohort. The use of 1994 as the treatment year is based on the legal mandate requiring all high schools to begin offering the course-at a minimum-to final-year students, making it unlikely that individuals who completed secondary school before 1994 received any exposure.    

Specifically, we estimate a difference-in-differences (DiD) model of the form:  
\begin{equation}
\label{baselineDiD}
\begin{multlined}
Y_{ics}=\phi_{c}+\gamma_{s}+\beta_{1}(\text{HighSchool}_{ics})+\beta_{2}\big(1{\text{[Age $\leq 16$ in 1994}]_{c}}\times 
\text{HighSchool}_{ics}\big)+\epsilon_{ics}
\end{multlined}
\end{equation}

where $Y_{ics}$ equals one if individual $i$ from cohort $c$ in survey year $s$ reports a given religious identity. Cohort ($\phi_c$) and survey-year ($\gamma_s$) fixed effects control for time-invariant cohort differences and common shocks. The interaction coefficient $\beta_2$ captures the causal effect of curriculum exposure.

All models also include region fixed effects. We additionally control for individual-level characteristics such as gender and ethnicity. Standard errors are clustered at the region level.  Given the small number of clusters, we report wild bootstrap confidence intervals following \citet{Cameronetalt2008}. In alternative specifications, we include interactions between cohort and region, high school and region, and survey year and region fixed effects, as well as controls for gender and ethnic minority. 

\medskip
The key identification assumption is that, absent the reform, treated and untreated cohorts would have followed parallel trends in religious identification. We address this in two ways. First, we restrict the sample to narrow age windows around the cutoff (e.g., [-3, +3]) to improve comparability. Second, we run placebo tests with pre-1994 cohorts, expecting no systematic treatment--control differences before exposure.\footnote{These design choices directly address concerns that cohort composition among high school completers may have shifted around the reform period. By focusing on adjacent cohorts, we minimize changes in observable and unobservable characteristics across treated and untreated groups, and the absence of pre-reform effects in placebo tests provides evidence against differential selection into high school completion driving the results.}

\medskip
Several potential threats remain. One concern is measurement error in treatment assignment, since we cannot observe the exact grade in which individuals were enrolled in 1994. However, education statistics suggest that most students graduated at standard ages, which limits this risk. A second concern is heterogeneity in implementation: public schools were more likely to adopt the constitutional curriculum than private schools, which retained greater curricular autonomy. Since the majority of high school graduates in the 1990s attended public schools, we expect that most treated students were in fact exposed, though imperfect compliance could attenuate the estimated effects. Finally, we acknowledge that the post-1994 changes in religious identity may also have been influenced by contemporaneous forces, such as the rapid expansion of Protestant churches. Nevertheless, our DiD framework compares relative changes across cohorts within the same regions and of roughly the same ages, making it unlikely that supply-side religious competition alone can account for the results.

\section{Baseline Results}
\label{sec_results}

Table~\ref{table_efftrcatholic} reports the baseline estimates of Equation (\ref{baselineDiD}), where the outcome is Catholic self-identification. Because the vast majority of Colombians identify as Catholic, any meaningful change in religious affiliation is most likely to appear in this category. We therefore begin by focusing on this measure. The table presents results from three specifications: (i) without fixed effects or controls (column 1), (ii) with cohort, region, and survey-year fixed effects (column 2), and (iii) with interacted fixed effects and additional controls for gender and ethnicity (column 3).

Across all specifications, the coefficients are negative and statistically significant. Exposure to the 1991 constitutional curriculum reduced the probability of self-identifying as Catholic by about three percentage points. This is one of the paper’s central findings. It is consistent with the patterns in Figure~\ref{fig_evolreladh} and supports the interpretation that the 1991 Constitution—implemented through mandatory civic education courses—contributed to a shift away from Catholic dominance.

The estimates in Table~\ref{table_efftrcatholic} are robust across specifications and age-window definitions, and they align with the identifying assumption of parallel trends. In particular, we find no evidence of differential pre-trends between treated and untreated cohorts prior to 1994. Section~\ref{section_robustness} provides a detailed discussion of these checks. Figure~\ref{fig_efftrcatholicall} illustrates this pattern: coefficients are statistically indistinguishable from zero before 1994, followed by a sharp and significant decline in Catholic self-identification that year. The figure also shows no comparable effects for post-1994 graduates relative to similarly aged peers who did not complete high school—a point we return to below.

Table \ref{table_efftrelignocathath} examines whether the reform affected other religious identities or overall religiosity. We construct binary indicators for: (i) non-Catholic Christian groups, including Evangelical, Pentecostal, Jehovah’s Witnesses, and Mormons (Panel A); (ii) non-Christian religions, such as Eastern faiths, Judaism, and traditional religions (Panel B); and (iii) no religion, agnosticism, or atheism (Panel C). Table~\ref{table_distributionreligident} displays the distribution of respondents across these categories, alongside Catholics, who represent the overwhelming majority.

The estimates reveal three main findings. First, Panel A shows a positive effect of about 1.9 percentage points on identification with non-Catholic Christian groups, though this estimate is not statistically significant. Second, Panel B shows no discernible effect on non-Christian affiliation. Third, Panel C shows a 1.5 percentage point increase in reporting no religion, agnosticism, or atheism.

\begin{figure}[H]
\begin{center}
\caption{Effect of Exposure to Constitutional Courses on Non-Catholic Christian, Non-Christian, and Secular Identification}
\label{fig_effectallpost1994}
\vspace{-0.3cm}
\begin{subfigure}{0.5\textwidth}
\caption{Non-Catholic Christian}
\includegraphics[width=8cm,height=6cm]{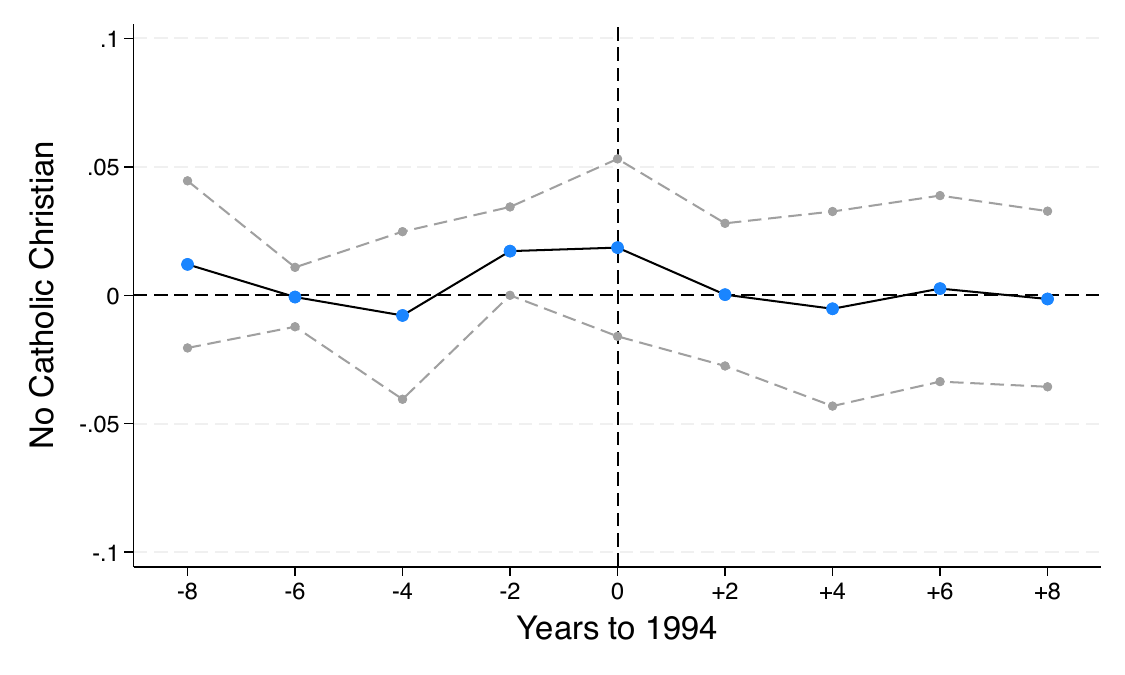}
\label{fig_efftrotherident_a}
\end{subfigure}\hspace*{\fill}
\begin{subfigure}{0.5\textwidth}
\caption{Non-Christian}
\includegraphics[width=8cm,height=6cm]{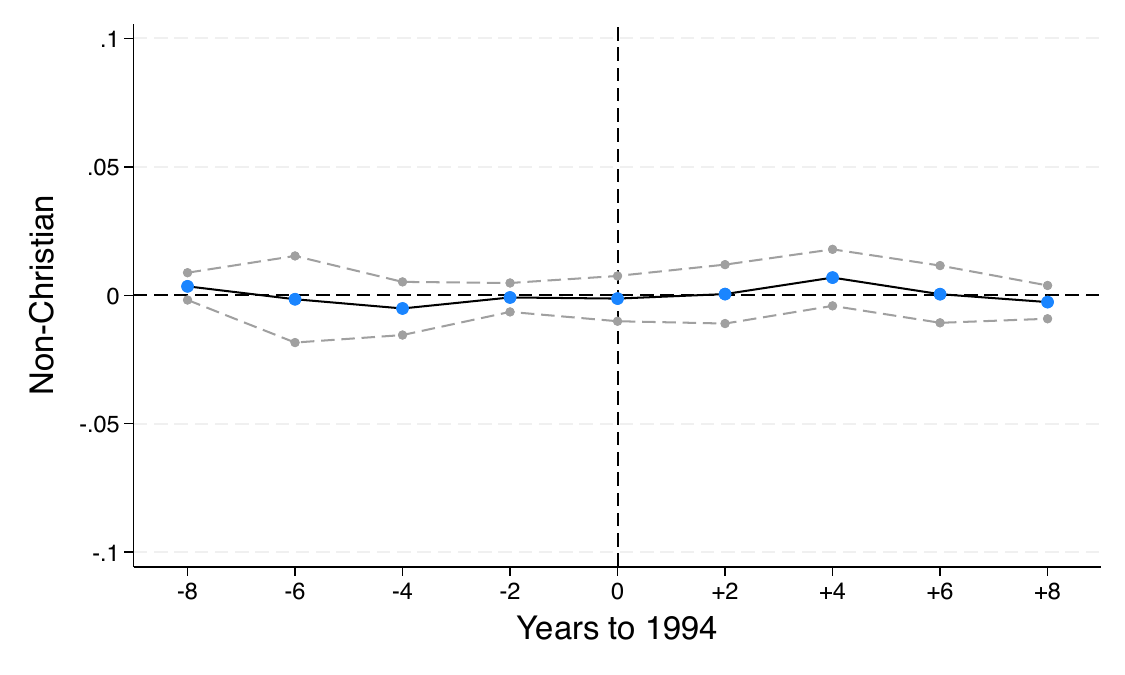}
\label{fig_efftrotherident_a}
\end{subfigure}
\begin{subfigure}{0.5\textwidth}
\caption{No religion or Atheist}
\includegraphics[width=8cm,height=6cm]{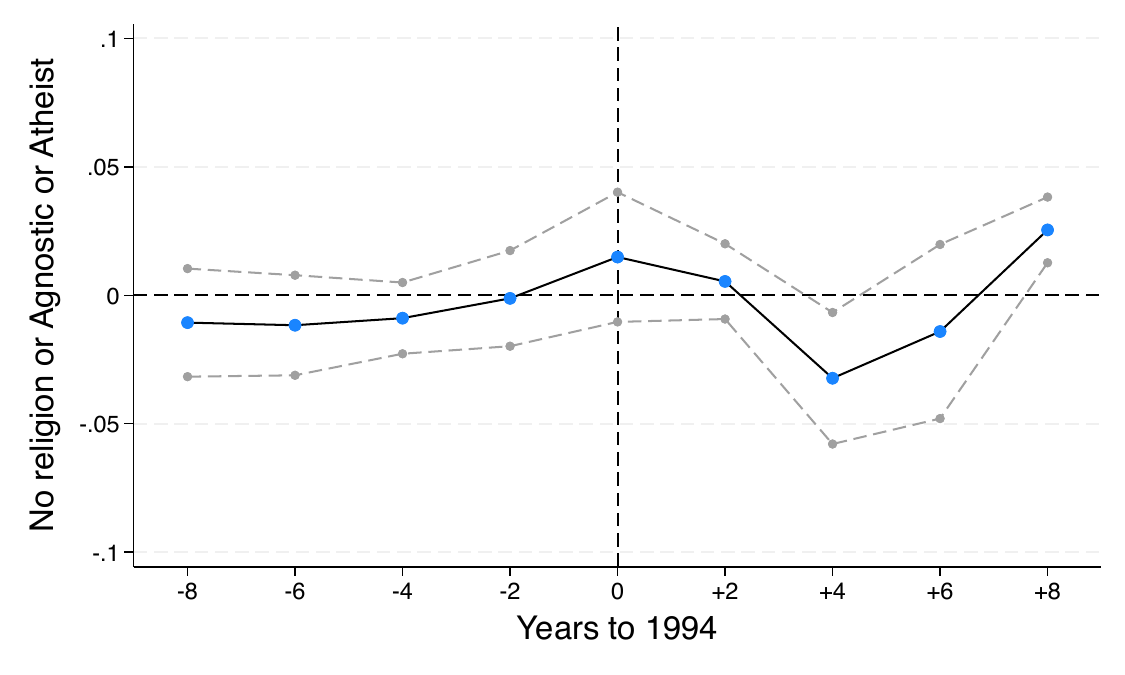}
\label{fig_efftrotherident_b}
\end{subfigure}
\begin{minipage}{15cm}
\footnotesize
\textbf{Note:} These figures plot the estimated effect of exposure to mandatory high school courses on the 1991 Constitution on identification with non-Catholic Christian, non-Christian, and secular affiliations. Each panel displays the coefficients from Equation~(\ref{baselineDiD}) using the specification reported in Table~\ref{table_efftrelignocathath} and an age window of $[-3, 3]$. 
The regression sample is drawn from the 2015, 2017, 2019, and 2021 waves of the Political Culture Survey conducted by DANE. The dependent variable is an indicator for self-identification as (a) a non-Catholic Christian denomination, (b) a non-Christian religion, or (c) no religion/agnostic/atheist. Points show the estimated coefficients by years relative to 1994, and the dashed lines depict the corresponding wild-bootstrap confidence intervals.
\end{minipage}
\end{center}
\end{figure}

Although the estimates are imprecise, they point to two complementary dynamics underlying the decline in Catholic affiliation: a shift toward secular identification and a possible reallocation toward non-Catholic Christian denominations. Section~\ref{sec_mechanisms} develops this interpretation and considers alternative explanations.\footnote{Can the previous results be extended to individuals who graduated in later years, who may have been exposed to the new curriculum not only during their final year but also in earlier grades? Assessing this question, however, is not straightforward. Students graduating after 1994 likely experienced more intensive exposure, but our data do not record the specific grade at which individuals left school. Consequently, we cannot determine with certainty whether those who dropped out before graduation avoided exposure altogether. Many may have received partial exposure if they left during grades in which the new curriculum had already been implemented, thereby blurring the distinction between treatment and control groups and complicating causal inference. With these caveats in mind, we re-estimate our models using alternative graduation cutoffs—1996, 1998, 2000, 2002, and 2004—focusing on the same outcome variables. Figure~\ref{fig_effectallpost1994} and Appendix Table~\ref{tab_effectallpost1994} summarize the results. Across all specifications, we find no effects for the 1996 graduates. For later cohorts, the evidence becomes more fragmentary, with most estimates statistically insignificant. We observe a temporary decline in the probability of reporting no religious or atheist affiliation among 1998 graduates, followed by increases in this outcome for the 2002 and 2004 cohorts. For the 2004 graduates, we also find a decline in Catholic self-identification. Taken together, the evidence suggests that the effects were concentrated among the transitional cohorts who experienced the reform as a visible institutional shift, while for post-1994 graduates, the estimates show scattered but inconsistent patterns.}

\section{Mechanisms}\label{sec_mechanisms}

How did exposure to courses on the 1991 Constitution lead to a decline in the probability of self-identifying as Catholic? In particular, which religious—or non-religious—identities did individuals adopt instead, and what factors may have driven this specific change? In this section, we address these questions and consider alternative explanations.

\subsection{Erosion of Catholic Monopoly and Legitimation of Pluralism}

Table~\ref{table_efftrelignocathath}, presented at the end of the previous section, suggests that the decline in Catholic identification may have been accompanied by a reconfiguration of religious identities. The estimates point to possible increases in both non-Catholic Christian affiliation and non-religious identification, although they remain noisy and not, on their own, conclusive. To deepen this initial insight, we now present additional evidence showing that the underlying dynamic does occur, but not uniformly across Colombia. Specifically, the evidence supports the following interpretation: civic education on the 1991 Constitution emphasized principles of religious pluralism, thereby weakening Catholicism’s symbolic monopoly and legitimizing alternative identities. By embedding pluralism into high school instruction during a formative stage of identity development, the reform made secular identification a more acceptable choice—particularly in regions where Catholic dominance had historically limited the presence of other denominations. Conversely, in regions where the Catholic Church, despite being the most established religion nationwide, was institutionally relatively weaker and religious competition stronger, the same curricular emphasis on pluralism likely interacted with existing diversity, reinforcing openness to Evangelical and Protestant identities.

To empirically examine this explanation, we begin by noting that Colombia is highly heterogeneous, and much of this variation is spatially patterned. A key dimension of interest concerns the presence and strength of religious institutions, especially the Catholic Church, which historically shaped local religious life. Our hypothesis is that this uneven spatial distribution is central to understanding how religious identities evolved after the curriculum reform. As shown in the previous section, exposure to the courses reduced Catholic identification on average, but the specific form this change took likely depended on the pre-existing religious landscape and institutional context.

A natural starting point is the distinction between Colombia’s center and periphery, which roughly corresponds to the Andean highlands (center) and the coastal Atlantic and Pacific regions (periphery).\footnote{Other peripheral regions include the Amazon and the Llanos. These areas are sparsely populated and not represented in our main survey sample, so we do not include them in the empirical analysis.} The 1886 Constitution and the 1887 Concordat entrenched Catholic authority over education, civil law, and public life, granting the Church extensive privileges and effective control over schooling, religious instruction, and teacher selection \citep{Arias2003, Beltran2013}. Throughout the twentieth century, this institutional arrangement sustained what historians have described as a religious monopoly, shaping national identity around \emph{nacional-catolicismo} and limiting the legitimacy of alternative denominations \citep{Beltran2013}. The institutional backbone of this alliance—archdioceses, seminaries, elite schools, and Conservative party networks—was concentrated in Bogota and other Andean departments such as Antioquia and Boyacá, where state power, education, and Catholic authority were most deeply intertwined \citep[][p.~126]{Beltran2013}. This configuration made the Andean core the historical epicenter of Catholic institutional strength.\footnote{For example, the main Catholic universities and seminaries—such as San Bartolomé, Javeriana, Santo Tomás and Pontificia Bolivariana—were located in Bogota, Antioquia and Boyacá, forming the intellectual and clerical base of the national Church and shaping political and educational elites \cite[][pp.~125–127]{Beltran2013}.}

By contrast, the Caribbean and Pacific regions have long exhibited weaker Catholic dominance and greater religious diversity. In the Caribbean, Protestant influences emerged early, with the Baptist Church well established by the mid-nineteenth century, and coastal mobility helped diffuse these networks \citep[][pp.~54–56]{Beltran2013}. More broadly, these regions were historically more peripheral to state power—less urbanized, with weaker state capacity and more fluid settlement patterns—which created a more open religious landscape. New religious movements grew most rapidly in places where both the Catholic Church and the State were institutionally weak, especially in colonization zones and frontier areas, often taking on community functions that Catholic parishes fulfilled elsewhere \citep[][pp.~49, 78–79, 170]{Beltran2013}.\footnote{\citet[][pp.~49, 78–79, 170]{Beltran2013} emphasizes that Protestant and Pentecostal movements flourished primarily in colonization frontiers, Indigenous territories, and Afro-descendant regions where Catholic and state institutions were weakest. These new churches often substituted for state and Catholic functions, offering social regulation, moral order, and community organization. The 1908 arrival of the Unión Misionera Evangélica in the southwest, followed by later missions in coastal and frontier nodes such as Cartagena, Santa Marta, Cali, and Pasto, exemplifies this pattern.}

\begin{figure}[H]
\begin{center}
\caption{Partial Correlation: Catholic Churches and Location in Central Regions}
\label{fig_distributionreligident6cat}
\vspace{-0.3cm}
\includegraphics[width=14cm,height=8cm]{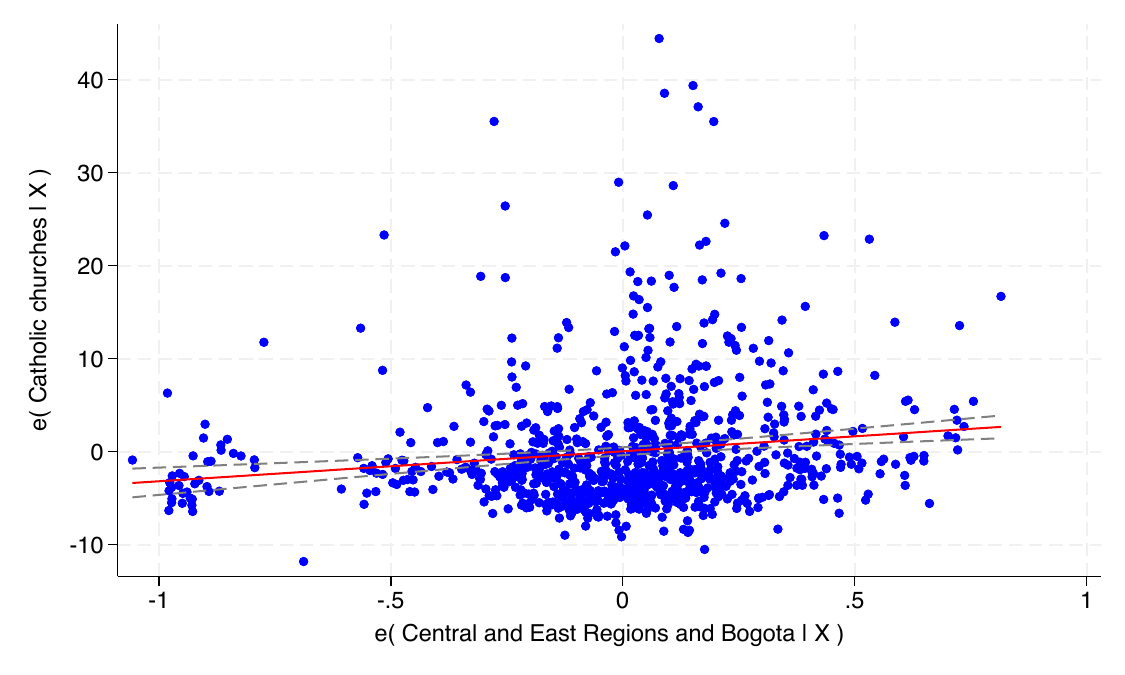}
 \begin{minipage}{14cm} \scriptsize\textbf{Note}: This figure presents a partial correlation plot illustrating the relationship between the number of Catholic churches per 100,000 inhabitants in 1995 and an indicator for whether the municipality belongs to the Central Regions (Central and Eastern regions and Bogota). The partial correlation is obtained from a regression that controls for the natural logarithm of total population (1995), municipal area, altitude, distance to the departmental capital, distance to Bogota, the share of rural population (1995), the Living Conditions Index (1995), and total municipal expenditure per capita (1995). Observations in the bottom and top 1\% of the dependent variable are excluded. Coef. = 3.22 (SE = 0.83). 
\end{minipage}
\end{center}
\end{figure}

Is the empirical evidence consistent with this regional contrast rooted in distinct historical trajectories? Figure~\ref{fig_distributionreligident6cat} presents the partial correlation between the municipal number of Catholic churches per 100{,}000 inhabitants in 1995—the only year for which such information is available- and a binary indicator for municipality located in the Andean regions.\footnote{We use a dataset from \textit{Fundación Social} containing 1995 municipal counts of Catholic churches and construct a per 100{,}000 measure. Although these data correspond to 1995—just after the 1994 introduction of the constitutional curriculum—church infrastructure is a slow-moving stock, making this year a reasonable approximation of pre-existing regional differences not mechanically driven by the reform. We estimate cross-sectional regressions on a binary indicator equal to one for municipalities located in Bogota, Central, and Eastern regions (Colombia’s Andean core). The partial correlation controls for $\ln(\text{population})$, municipal revenue and expenditure, and a composite living-conditions index, and includes department fixed effects. Standard errors are clustered at the municipal level.} The figure reveals a clear positive relationship: municipalities with more Catholic churches per capita are disproportionately located in the Andean core, consistent with the idea that Catholicism historically maintained a stronger institutional presence in these regions. Columns (1) and (2) of Appendix Table~\ref{tab_corrcathethniregions} further confirm that this relationship is statistically significant and robust to alternative controls.\footnote{An important related issue is whether it is also possible to empirically test the historically greater presence of non-Catholic Christian churches in the regions we classify as peripheral. This test is considerably more difficult—and, given current data, not feasible—because reliable pre-1996 information on non-Catholic congregations is largely unavailable. This scarcity is plausibly due to heterogeneous record-keeping practices and the prevalence of informal or unregistered congregations; indeed, a formal national registry for these organizations did not begin to exist until 1996.}

If the center–periphery contrast discussed above is plausible, a natural next step is to ask whether this uneven spatial distribution of religious institutions helps explain the regionally differentiated evolution of religious identities in Colombia—beyond the average decline in Catholic identification documented in Section~\ref{sec_results}. Using the baseline sample underlying our main estimates, we distinguish between two broad types of regions that align with the qualitative patterns previously described.\footnote{These regions are: (i) the Pacific and Atlantic regions, which include the departments of Atlántico, Bolívar, Cesar, Córdoba, La Guajira, Magdalena, Sucre, Chocó, Cauca, Nariño, and Valle del Cauca; and (ii) the Central and Eastern regions together with Bogota, which include Antioquia, Boyacá, Caldas, Caquetá, Cundinamarca, Huila, Meta, Norte de Santander, Quindío, Risaralda, Santander, and Tolima. Bogota is grouped with these regions due to its central institutional role. Region-level identifiers are the most disaggregated units available in the survey; ideally, finer geographic variation would be preferable.} Columns (1)–(4) of Table~\ref{tab_effectallregion} show that three patterns emerge: first, Catholic identification declines in both types of regions; second, in the Central and Eastern regions plus Bogota, the decline in Catholic identification is  associated with an increase in secular identification but not with changes in non-Catholic Christian affiliation; and third, in the Pacific and Atlantic regions, this decline is accompanied instead by an increase in identification with non-Catholic Christian denominations but no significant change in secular identification. Taken together, these results provide clear support for our hypothesis that the reform produced a heterogeneous reconfiguration of religious identities across Colombia.\footnote{A potential concern with these heterogeneity exercises is the use of relatively coarse geographic units and the possibility of selective migration across regions. Unfortunately, the survey does not contain information on respondents’ migration histories (e.g., duration of residence or previous location). As a partial check, we re-estimate the main specifications excluding Bogota—the region with the highest internal migration rates. The results, reported in Table~\ref{tab_effectallregionnobogota}, remain qualitatively unchanged, suggesting that migration is unlikely to drive the documented regional patterns.}

A second empirical exercise that provides additional evidence for our hypothesis examines whether an additional dimension closely linked to the identity shifts documented above also evolved in a manner consistent with our proposed mechanism: religious attendance.\footnote{Religious attendance is obtained from the same \textit{Political Culture Survey} used in the baseline results. The relevant question asks: “Cada cuánto usted asiste a reuniones de las siguientes organizaciones de participación voluntaria: Iglesias, organizaciones o grupos religiosos. 1. Una vez a la semana; 2. Una o dos veces al mes; 3. Una o dos veces al año; 4. Nunca.” (“How often do you attend meetings of the following voluntary organizations: churches, organizations, or religious groups? 1. Once a week; 2. Once or twice a month; 3. Once or twice a year; 4. Never.”) We define a dummy equal to one if the respondent selects 1 or 2, and zero otherwise.} Column (5) of Table~\ref{tab_effectallregion} reports the results: in the Pacific and Atlantic regions, religious attendance increases (Panel B, column (5)), while in the Central and Eastern regions it decreases (Panel A, column (5)). This pattern aligns with the previous findings. In peripheral regions, the decline in Catholic identification appears to be absorbed primarily by non-Catholic Christian groups—where religious practice is typically more intensive; whereas in central regions, the shift toward secular identification is accompanied by lower attendance.

\medskip
A central issue within the center–periphery distinction concerns another dimension closely correlated with this geographic divide—and one that may also be crucial for explaining our main results: ethnic identification. Historical evidence shows marked ethnic differences between the center–periphery areas we previously defined. Colombia’s central regions are comparatively more ethnically homogeneous, with the vast majority of the population identifying as White or Mestizo, whereas the peripheral regions are more heterogeneous and contain a substantially higher concentration of ethnic minorities. These contrasts align with the empirical patterns presented in Figure~\ref{fig_distributionreligident6tehnicminor} and Columns (3) and (4) of Appendix Table~\ref{tab_corrcathethniregions}. Both the figure and the table display the correlation between the share of ethnic minorities in Colombian municipalities in 1993—the closest pre-reform year for which data are available\footnote{The data come from the 1993 Population Census and classify Afro-descendant and Indigenous populations as ethnic minorities. Our measure of ethnic minorities is therefore based solely on these two groups.}—and a binary indicator for municipalities located in the Andean regions. The figure and table reveal a negative and statistically significant relationship, confirming that ethnic minorities are more prevalent in Colombia’s peripheral regions.

Given this ethnic segregation, it is not immediately clear whether the heterogeneous regional effects documented above operate similarly across ethnic groups or whether they may instead be confounded by them. What, then, is the role of ethnic identification in the mechanism we have proposed to explain the effect of exposure to the 1991 constitutional curriculum on religious identification?

To explore this question, we draw on the self-reported ethnic identification available in our baseline data and assess whether the relationship between exposure to the 1991 constitutional curriculum and religious identity differs across two broad groups that together represent a substantial share of our sample: individuals who identify with an ethnic minority (16.6\%) and those who do not (83.4\%).\footnote{The survey allows respondents to identify as Afro-Colombian (10.5\%), Indigenous (5.8\%),  Palenquero (0.16\%), Raizal (0.05\%), Rom (0.04\%), or none of these (83.45\%). We classify as “ethnic minority” those who select any of the minority categories, and as “non-minority” those who report no affiliation with them.}
Table~\ref{tab_effectallethnicityregion} reports the results of this analysis. Panels A and B present estimates from our baseline specification for each group separately. Three main findings emerge.
First, Catholic identification declines among both groups (column (1)); for the ethnic minority group, however, estimates are substantially noisier due to its smaller sample size.
Second, there are no statistically significant differences in the change in identification with non-Catholic Christian groups or with no religion/atheism (columns (2) and (4)).
Third, while ethnic minorities exhibit an increase in religious attendance, non-minorities show a decrease (column (5)).

Are these results consistent with those reported in Table~\ref{tab_effectallregion} and with our preferred interpretation? To address this question, Panels C–F of Table~\ref{tab_effectallethnicityregion} combine the regional and ethnic dimensions, yielding four sets of heterogeneous effects by region and minority status.
The results show the following main patterns.
First, Catholic identification declines across all subgroups (column (1)).
Second, non-Catholic Christian affiliation rises among non-minorities in peripheral regions (Panel D, column (2)) and among minorities in central regions (Panel E, column (2)).
Third, secular identification increases only in the central region—particularly among non-minorities (column (4)).
Fourth, religious attendance rises among ethnic minorities regardless of region (column (5)), but declines among non-minorities, especially in central areas (Panel F, column (5)).

Taken together, these results are consistent with our preferred hypothesis while revealing two additional patterns that constitute the paper’s second main contribution. First, the decline in Catholic identification  translated into new forms of religious identity involving more intensive practice, particularly in regions with weaker Catholic presence and stronger religious competition, and mainly among non-minority individuals. This is consistent with our previous explanation that non-Catholic churches in peripheral regions historically targeted Indigenous and Afro-descendant communities: if these minority groups were already less Catholic by 1994, the reform’s effect would have primarily operated among non-minorities living in more religiously competitive environments, where established non-Catholic denominations were viable alternatives.\footnote{An additional statistically significant result of non-negligible magnitude appears in Panel E, column (2) of Table~\ref{tab_effectallethnicityregion}: an increase in non-Catholic Christian affiliation among ethnic minorities in the country’s central regions. Although this differs from the pattern in Panel B, it remains broadly consistent with our preferred interpretation. In central regions, ethnic minorities are often recent or fragmented migrants with comparatively weak ancestral ties, typically concentrated in peripheral urban neighborhoods where non-Catholic Christian churches have expanded in recent decades \citep{Beltran2013}. However, because the number of observations in this subgroup is very small, we refrain from placing much emphasis on this finding.} Second, the decline in Catholicism also entailed a broader reduction in religiosity, but this pattern emerged mainly in areas with limited religious competition. In these regions—where both minority and non-minority individuals lacked viable alternative denominations—the weakening of Catholic dominance appears to have led not to denominational substitution, but to secularization and declining religious participation.\footnote{Disaggregating the effects by gender also reveals an interesting, though ultimately too noisy, pattern: declines in Catholic identification and increases in alternative affiliations appear for both men and women, but are plausibly stronger and more precisely estimated among men (Appendix Table~\ref{tab_effectallgender}). This is consistent with the notion that men—historically less embedded in Catholic ritual life and community networks—were more responsive to the legitimization of pluralism and secular options, whereas women’s deeper ties to Catholic institutions limited such shifts.}

\subsection{Alternative Explanations: Religiosity, Ideology and the Economy}

While the erosion of the Catholic monopoly and the legitimation of pluralism provide the most coherent explanation for our findings, we also consider alternative mechanisms.

A first alternative mechanism concerns an outcome examined in the previous subsection—religious attendance. One possibility is that exposure to courses on the 1991 Constitution may have made religious practice less intense or less salient, thereby contributing to the decline in Catholic identification. Column (5) of Table~\ref{tab_effectallethnicityregion}, already discussed above, directly evaluates this channel. The results show that among individuals who self-identify as ethnic minorities, the decline in Catholic identification is accompanied by an increase in religious attendance, effectively ruling out this interpretation. Panel A of Table~\ref{table_altermechanisms} complements this evidence: when estimating the effect of the new curriculum on religious attendance in the full sample (without distinguishing between ethnic groups), the coefficients are statistically insignificant. Thus, rather than weakening religious practice, the reform appears to have redirected it.

A second alternative mechanism is that exposure to civic courses may have shifted political ideology, making individuals less conservative and thereby less attached to Catholicism—since in Colombia, the Catholic Church has historically been associated with conservative positions across multiple dimensions \citep[see, for example,][]{Arias2003}. Panel B of Table~\ref{table_altermechanisms} examines this channel, using as the dependent variable respondents’ self-reported political ideology, measured on a 1–10 scale where higher values correspond to more right-wing positions.\footnote{Political ideology is obtained from the same \textit{Political Culture Survey} used in the baseline results. The relevant question is: “En una escala de 1 a 10, donde 1 significa izquierda y 10 significa derecha, ¿dónde se ubicaría usted en esta escala?” (“On a scale from 1 to 10, where 1 means left and 10 means right, where would you place yourself?”). We use this variable directly as reported in the survey.} The estimates are negative—consistent with a potential shift away from conservatism—but statistically indistinguishable from zero. This suggests that while ideological change may have played a limited role, it does not appear to be the primary mechanism driving our results.

A third possibility is that economic factors may have facilitated switching. Exposure to the new curriculum could, in principle, have influenced labor market participation and, through this channel, shaped attitudes toward religious pluralism or non-affiliation. Panel C of Table~\ref{table_altermechanisms} evaluates this channel, using as the dependent variable a dummy equal to one if respondents reported being employed or seeking employment.\footnote{Labor market participation is obtained from the same \textit{Political Culture Survey} used in the baseline results. The survey question is: “Actualmente, ¿usted trabaja o está buscando trabajo?” (“Are you currently working or looking for work?”). Respondents are coded as 1 if they reported working or actively seeking work.} We find no evidence that exposure to the curriculum affected labor force participation.

Taken together, these results reinforce our preferred mechanism. The evidence suggests that the curriculum legitimized denominational pluralism within Christianity, weakened Catholic dominance, and facilitated switching both toward Evangelical and Protestant groups and toward secular identification—without altering individuals’ political ideology or their participation in the labor market.

\section{Robustness Checks}
\label{section_robustness}

In this section, we assess the robustness of our findings through complementary exercises designed to address potential threats to identification.

\begin{figure}[H]
\begin{center}
\caption{Effect of Exposure to Courses on the 1991 Constitution on Catholic Identification}
\label{fig_efftrcatholicall}
\vspace{-0.3cm}
\includegraphics[width=14cm,height=9cm]{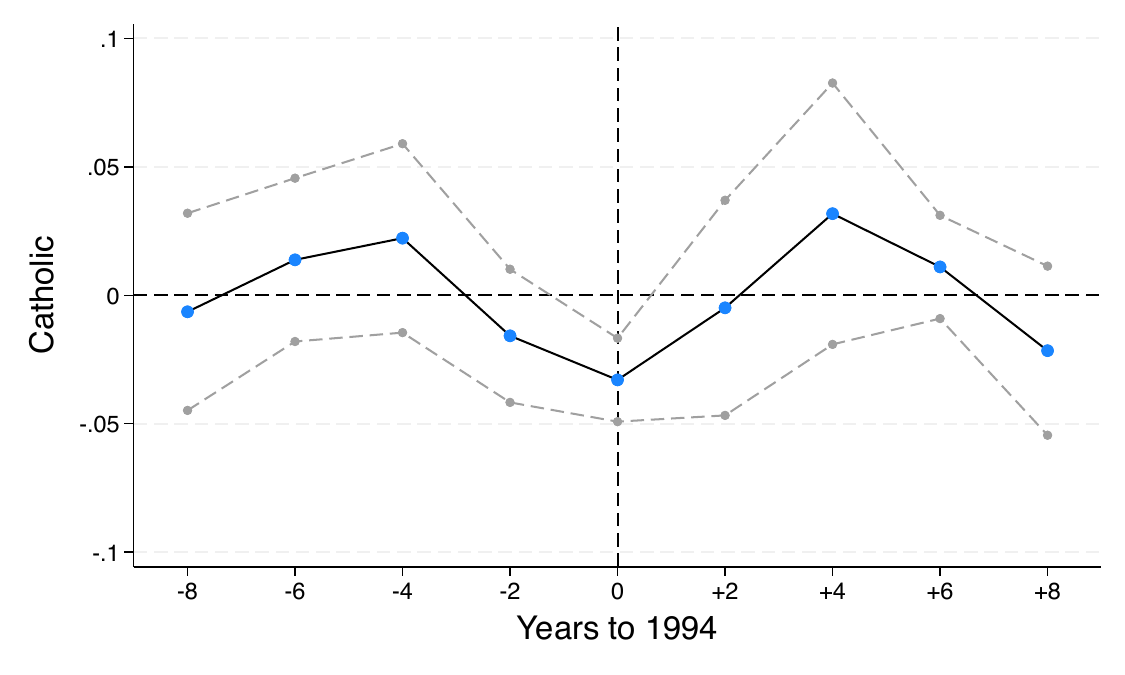}
 \begin{minipage}{14cm} \footnotesize\textbf{Note}: This figure plots the estimated effect of exposure to mandatory high school courses on the 1991 Constitution on Catholic self-identification, using the baseline specification of Equation (\ref{baselineDiD}) reported in column (3) of Table \ref{table_efftrcatholic}, but extending the estimation to individuals who completed high school in years both prior and subsequent to 1994. Points depict the estimated coefficients by years relative to 1994, and the dashed lines display the corresponding confidence intervals.
\end{minipage}
\end{center}
\end{figure}

A first concern is that the results may reflect pre-existing differences in religious identity between high school graduates and non-graduates, rather than the effect of the curriculum. Figures~\ref{fig_efftrcatholicall} and \ref{fig_effectallpost1994} already speak to this issue by re-estimating Equation~(\ref{baselineDiD}) using pre-1994 data, showing that graduation had no effect on identification with any religious denomination—or on non-religiosity or atheism—prior to the reform. We now present and discuss more comprehensive results along these lines. Table~\ref{tab_efftrcatholi_before94_w234} reports the baseline estimates (column (2); those displayed in Figures~\ref{fig_efftrcatholicall} and \ref{fig_distributionreligident6cat}), alongside estimates using a narrower age window ([-2,2], column (1)) and a wider window ([-4,4], column (3)). All coefficients are statistically indistinguishable from zero. Analogous results for the other outcomes are shown in Appendix Tables~\ref{tab_efftrnocatchriall_before94_w234}, \ref{tab_efftrnochristian_before94_w234}, and \ref{tab_efftrnoreligion_before94_w234}. Taken together—and with one notable exception—these findings provide no evidence of pre-existing differences between treated and untreated individuals, thereby strengthening the credibility of the parallel trends assumption.\footnote{One important exception arises for identification with non-Catholic Christian denominations in the age window $[-3, 3]$ for the year 1992, as shown in column (2) of Table~\ref{tab_efftrnoreligion_before94_w234}. In this case, we observe a positive and statistically significant coefficient of similar magnitude to the one estimated for the same outcome and age window in 1994. This effect does not appear for other age windows or other years (e.g., the effect for 1990 is zero). Although we do not have a definitive explanation for this pattern, one possibility is that by 1992 discussions about expanding religious pluralism in schools had already begun, in anticipation of the new cultural and institutional landscape introduced by the 1991 Constitution. If so, individuals graduating in 1992 may have been partially “treated,” even without formal exposure to the new courses. Consistent with this interpretation, the effect for 1990 is zero, and graduates from that year clearly could not have been exposed to such early adjustments. Hence, 1990 graduates constitute a cleaner placebo group than those from 1992. Regarding the other outcomes, we observe a few additional statistically significant results, but these are scattered, very small in magnitude, and opposite in sign to the main findings. Specifically: (i) a 0.5-percentage-point increase in identification with non-Christian groups among the 1986 graduates under the $[-3, 3]$ and $[-4,4]$ age windows (Appendix Table~\ref{tab_efftrnochristian_before94_w234}); and (ii) a 1.2-percentage-point decline in identification with no religion or atheism among the 1986 graduates for the $[-2,2]$ and $[-3,3]$ windows, and among the 1990 graduates for the $[-4,4]$ window (Appendix Table~\ref{tab_efftrnoreligion_before94_w234}).}

A second concern is the sensitivity of the main results to the choice of age window ([-3,3]) used in the baseline analysis. As discussed earlier, this bandwidth balances two considerations: keeping the window narrow enough to ensure plausibility of similar outcome trends between treated and control groups, while wide enough to capture individuals with sufficient exposure to the reform. Figure \ref{fig_efftrcatholi_agewin} examines the sensitivity of our main outcome (Catholic identification) to alternative age windows ranging from [-2,2] to [-8,8]. The figure shows that the estimated coefficients are of similar magnitude for windows close to the baseline ([-2,2] and [-4,4]) but vary for wider windows, likely because the identification assumptions are less plausible. As for the noisiness of the estimates, we note that results for the [-2,2] window are systematically less precise, which is expected given the smaller sample size and potential measurement error. A similar phenomenon is obtained for the other outcomes, as shown in Figure \ref{fig_efftrotherall_agewin}. Table \ref{tab_efftrotherall_agewin} presents the estimates for all the outcomes.

Taken together, the robustness checks reinforce the credibility of our findings. The absence of pre-trends and the stability of results across small bandwidths supports the conclusion that civic education on the 1991 Constitution led exposed individuals to view both non-Catholic affiliations and non-religious identification as more legitimate and socially acceptable.

\section{Conclusion}
\label{sec_conclusion}

This paper examines the effects of civic education on religious identification in Colombia following the 1991 Constitution. Exploiting cohort-based variation in exposure to mandatory constitutional courses introduced in high schools in 1994, and using repeated cross-sections from the nationally representative Political Culture Survey, we implement a difference-in-differences design. The results show that exposure to the constitutional curriculum reduced the probability of self-identifying as Catholic by about three percentage points, with no evidence of pre-existing differences or differential pre-trends across cohorts.

This decline does not reflect a generalized weakening of religiosity but rather a reallocation of religious identities. We document two complementary patterns: increased identification with non-Catholic Christian denominations—particularly Evangelical and Protestant groups—and a rise in secular identification. Together, these findings suggest that the reform weakened Catholicism’s symbolic monopoly and lowered the social and normative costs of alternative affiliations, including non-affiliation, without reducing religious engagement overall.

These patterns align with perspectives that emphasize the responsiveness of religious affiliation to institutional arrangements and symbolic legitimacy, and they complement recent evidence highlighting religious adaptation and diversification rather than uniform secular decline in emerging and developing countries. In this setting, state-led educational reforms play a central role by shaping the relative legitimacy of competing religious and secular identities.

Several limitations remain. Because outcomes are observed only from 2015 onward, we cannot trace short-run adjustment dynamics. Moreover, while the evidence supports a pluralism-based mechanism, we cannot fully disentangle the effects of civic education from contemporaneous developments such as the rapid expansion of Protestant churches during the 1990s. Future research could draw on alternative data sources or qualitative evidence to examine these processes more directly. 

Overall, the paper provides evidence that civic education can have persistent effects on cultural identities. By institutionalizing pluralism through schooling, the Colombian state contributed to a lasting reconfiguration of religious identification, weakening Catholic dominance while opening space—albeit unevenly—for both denominational diversity and secular affiliation.

\clearpage
\newpage

\bibliographystyle{ecca}
\bibliography{bib}

\hbox {} \newpage

\section*{Figures and Tables}


\begin{table}[H]
\begin{center}
{
\renewcommand{\arraystretch}{1.1}
\setlength{\tabcolsep}{12pt}
\caption {Distribution of Religious Identification Across All Groups}  \label{table_distributionreligident}
\vspace{-0.3cm}
\small
\centering  
\begin{tabular}{lcc}
\hline\hline \addlinespace[0.1cm]
    & (1)& (2)\\\addlinespace[0.1cm]\cmidrule[0.2pt](l){2-3}\addlinespace[0.1cm]
        & N& \%\\\addlinespace[0.1cm]\cmidrule[0.2pt](l){1-3}\addlinespace[0.1cm]
\textbf{Catholic}&      \textbf{123127}&      \textbf {81.60}\\
\textbf{Non-Catholic Christian}&      \textbf{17246}&      \textbf{11.43}\\
\hspace{3mm}Evangelical or Pentecostal&        9999&        6.63\\
\hspace{3mm}Protestant  &        6100&        4.04\\
\hspace{3mm}Jehovah's Witness&        1041&        0.69\\
\hspace{3mm}Mormon Church&         106&        0.07\\
\textbf{Non-Christian}&      \textbf{1548}&      \textbf{1.02}\\
\hspace{3mm}Eastern non-Christian&        1470&        0.97\\
\hspace{3mm}Jewish      &          48&        0.03\\
\hspace{3mm}Traditional religions&          30&        0.02\\
\textbf{No religion or Atheist}&      \textbf{8966}&      \textbf{5.94}\\
\hspace{3mm}No religion &        8437&        5.59\\
\hspace{3mm}Agnostic or Atheist&         529&        0.35\\
\textbf{Total}      &     \textbf{150887}&      \textbf{100.00}\\

\addlinespace[0.1cm]\hline\hline\addlinespace[0.1cm]
\multicolumn{3}{p{9cm}}{\scriptsize{\textbf{Note:} This table reports the distribution of religious identification in the sample of individuals used for the main results. The data come from DANE’s Political Culture Survey (waves 2015, 2017, 2019, and 2021). The exact survey question is: “Si usted es de alguna religión, ¿puede decirme cuál es su religión?” (“If you belong to a religion, could you tell me which religion it is?”). Response options included Catholic; Protestant (traditional or non-Evangelical Protestant); non-Christian Eastern religions; none; Evangelical or Pentecostal; Church of Jesus Christ of Latter-day Saints; traditional religions; Jewish; agnostic or atheist; and Jehovah’s Witness.}} 
\end{tabular}
}
\end{center}
\end{table}


\begin{table}[H]
\begin{center}
{
\renewcommand{\arraystretch}{0.9}
\setlength{\tabcolsep}{10pt}
\caption {Effect of Exposure to Courses on the 1991 Constitution on Catholic Identification}  \label{table_efftrcatholic}
\vspace{-0.3cm}
\footnotesize
\centering  
\begin{tabular}{lccc}
\hline\hline \addlinespace[0.15cm]
    & (1)& (2)& (3)\\\addlinespace[0.12cm]\cmidrule[0.2pt](l){2-4}\addlinespace[0.12cm]
                      &\multicolumn{3}{c}{Dependent Variable: Catholic Identification}\\\addlinespace[0.10cm]\hline
                \addlinespace[0.15cm]         
(Age $\leq$ 16 in 1994) $\times$ (High school) & -0.028 & -0.028 & -0.033 \\
  & (0.006)*** & (0.006)*** & (0.006)*** \\
  & [-0.042, 0.007]* & [-0.039, 0.006]* & [-0.046, -0.021]*** \\
R-sq  & 0.001 & 0.012 & 0.021 \\
Observations  & 19,875 & 19,875 & 19,875 \\

    \addlinespace[0.15cm]\hline    \addlinespace[0.15cm]
\multicolumn{1}{l}{Cohort, region \& year fixed effects}     &  N& Y& Y \\
\multicolumn{1}{l}{Interacted fixed effects and controls}    &  N& N& Y \\
\addlinespace[0.15cm]\hline\hline\addlinespace[0.1cm]
\multicolumn{4}{p{14.5cm}}{\scriptsize{
\textbf{Note:} This table presents the estimated effect of exposure to mandatory high school courses on the 1991 Constitution on Catholic self-identification. All columns report estimates of Equation (\ref{baselineDiD}) using an age window of $[-3, 3]$. 
The regression sample is drawn from the 2015, 2017, 2019, and 2021 waves of the Political Culture Survey conducted by DANE. 
The dependent variable in all columns is an indicator equal to 1 for respondents who selected ``Catholic'' in response to the question ``If you belong to a religion, could you tell me which religion it is?''. 
Specifications in column (2) include fixed effects for age cohort, region, and survey year. 
Column (3) additionally includes dummy variables for female and ethnic-minority status, as well as cohort-by-region, high school-by-region, and survey-year-by-region fixed effects. 
Robust standard errors clustered at the region level are reported in parentheses, and wild-bootstrap confidence intervals are shown in square brackets. 
*, **, and *** indicate statistical significance at the 10\%, 5\%, and 1\% levels, respectively.
}}
\end{tabular}
}
\end{center}
\end{table}


\begin{table}[H]
\begin{center}
{
\renewcommand{\arraystretch}{0.8}
\setlength{\tabcolsep}{15pt}
\caption{Effect of Exposure to Constitutional Courses on Non-Catholic Christian, Non-Christian, and Secular Identification} 
\vspace{-0.3cm}
\label{table_efftrelignocathath}
\footnotesize
\centering  
\begin{tabular}{lccc}
\hline\hline \addlinespace[0.15cm]
    & (1)& (2)& (3)\\\addlinespace[0.12cm]\hline\addlinespace[0.12cm]
      \multicolumn{1}{l}{\emph{\underline{Panel A}:  }}      & \multicolumn{3}{c}{Dependent Variable: Non-Catholic Christian}\\\addlinespace[0.1cm]\cmidrule[0.2pt](l){2-4}\addlinespace[0.1cm]               
(Age $\leq$ 16 in 1994) $\times$ (High school) & 0.015 & 0.015 & 0.019 \\
  & (0.011) & (0.011) & (0.012) \\
  & [-0.022, 0.052] & [-0.020, 0.050] & [-0.014, 0.051] \\
R-sq  & 0.000 & 0.009 & 0.017 \\
Observations  & 19,875 & 19,875 & 19,875 \\

\addlinespace[0.15cm]\hline\addlinespace[0.15cm]

      \multicolumn{1}{l}{\emph{\underline{Panel B}:  }}      & \multicolumn{3}{c}{Dependent Variable: Non-Christian}\\\addlinespace[0.1cm]\cmidrule[0.2pt](l){2-4}\addlinespace[0.1cm]               
(Age $\leq$ 16 in 1994) $\times$ (High school) & -0.001 & -0.001 & -0.001 \\
  & (0.003) & (0.003) & (0.003) \\
  & [-0.006, 0.007] & [-0.007, 0.007] & [-0.006, 0.007] \\
R-sq  & 0.000 & 0.001 & 0.005 \\
Observations  & 19,875 & 19,875 & 19,875 \\

\addlinespace[0.15cm]\hline\addlinespace[0.15cm]
      \multicolumn{1}{l}{\emph{\underline{Panel C}:  }}      & \multicolumn{3}{c}{Dependent Variable:  No Religion or Atheist}\\\addlinespace[0.1cm]\cmidrule[0.2pt](l){2-4}\addlinespace[0.1cm]               
(Age $\leq$ 16 in 1994) $\times$ (High school) & 0.014 & 0.014 & 0.015 \\
  & (0.010) & (0.009) & (0.009) \\
  & [-0.004, 0.045] & [-0.003, 0.045]* & [-0.001, 0.045]* \\
R-sq  & 0.001 & 0.007 & 0.014 \\
Observations  & 19,875 & 19,875 & 19,875 \\

    \addlinespace[0.15cm]\hline    \addlinespace[0.15cm]
\multicolumn{1}{l}{Cohort, region \& year fixed effects}     &  N& Y& Y \\
\multicolumn{1}{l}{Interacted fixed effects}    &  N& N& Y \\
\addlinespace[0.15cm]\hline\hline\addlinespace[0.15cm]
\multicolumn{4}{p{15.1cm}}{\scriptsize{
\textbf{Note:} This table presents the estimated effect of exposure to mandatory high school courses on the 1991
Constitution on identification with non-Catholic Christian, non-Christian, and secular affiliations. All columns report estimates of Equation (\ref{baselineDiD}) using an age window of $[-3, 3]$. 
The regression sample is drawn from the 2015, 2017, 2019, and 2021 waves of the Political Culture Survey conducted by DANE. The dependent variable in Panel A is identification with non-Catholic Christian denominations (including Evangelical, Pentecostal, Jehovah’s Witness, and Mormon groups). The dependent variable in Panel B is identification with non-Christian religions (including Eastern religions, Judaism, and traditional religions). The dependent variable in Panel C is identification with no religion/agnostic/atheist. Specifications in column (2) include fixed effects for age cohort, region, and survey year. Column (3) additionally includes dummy variables for female and ethnic-minority status, as well as cohort-by-region, high school-by-region, and survey-year-by-region fixed effects. Robust standard errors, clustered at the region level, are reported in parentheses, and wild-bootstrap confidence intervals are shown in square brackets. 
*, **, and *** indicate statistical significance at the 10\%, 5\%, and 1\% levels, respectively.
}}

\end{tabular}
}
\end{center}
\end{table}

\begin{table}[H]
\vspace{3cm}
\begin{center}
{
\renewcommand{\arraystretch}{0.8}
\setlength{\tabcolsep}{3pt}
\caption {Effect of Exposure to 1991 Constitution Courses on Religious Identification: Heterogeneous Effects by Region}  \label{tab_effectallregion}
\vspace{-0.3cm}
\scriptsize
 \begin{tabular}{lccccc}
\hline\hline \addlinespace[0.15cm]
    & (1)& (2)& (3)  & (4)& (5)\\\addlinespace[0.12cm]\cmidrule[0.2pt](l){2-6}\addlinespace[0.12cm]
& \multicolumn{5}{c}{Dependent Variable: Identification with the following groups}\\\addlinespace[0.15cm]\cmidrule[0.2pt](l){2-6} \addlinespace[0.10cm]   

&\multicolumn{1}{c}{Catholic}&\multicolumn{1}{c}{Non-Catholic}&\multicolumn{1}{c}{Non-Christian}&\multicolumn{1}{c}{No Religion}&\multicolumn{1}{c}{Religious}\\

&\multicolumn{1}{c}{}&\multicolumn{1}{c}{Christian}&\multicolumn{1}{c}{}&\multicolumn{1}{c}{or Atheist}&\multicolumn{1}{c}{attendance}\\

\addlinespace[0.15cm]\hline\addlinespace[0.15cm]

\multicolumn{1}{l}{\emph{\underline{Panel A}:  }}      & \multicolumn{5}{c}{Central and East regions and Bogota}\\\addlinespace[0.1cm]\cmidrule[0.2pt](l){2-6}\addlinespace[0.1cm]  
(Age $\leq$ 16 in 1994) $\times$ (High school) & -0.028 & 0.004 & -0.004 & 0.028 & -0.076 \\
  & (0.009)* & (0.016) & (0.002) & (0.010)* & (0.037) \\
  & [-0.041, -0.013]*** & [-0.016, 0.028] & [-0.006, -0.002]*** & [0.015, 0.042]*** & [-0.139, -0.033]*** \\
R-sq  & 0.007 & 0.007 & 0.001 & 0.007 & 0.019 \\
Observations  & 10,542 & 10,542 & 10,542 & 10,542 & 4,723 \\

\addlinespace[0.18cm]\hline\addlinespace[0.12cm]
\multicolumn{1}{l}{\emph{\underline{Panel B}:  }}      & \multicolumn{5}{c}{Pacific and Atlantic Regions}\\\addlinespace[0.1cm]\cmidrule[0.2pt](l){2-6}\addlinespace[0.1cm]  
(Age $\leq$ 16 in 1994) $\times$ (High school) & -0.035 & 0.030 & 0.002 & 0.002 & 0.028 \\
  & (0.012) & (0.019) & (0.006) & (0.001) & (0.011) \\
  & [-0.048, -0.022]*** & [0.013, 0.052]*** & [-0.005, 0.008] & [-0.001, 0.001] & [0.015, 0.041]*** \\
R-sq  & 0.009 & 0.005 & 0.002 & 0.009 & 0.056 \\
Observations  & 9,333 & 9,333 & 9,333 & 9,333 & 3,486 \\

\addlinespace[0.15cm]\hline\hline\addlinespace[0.15cm]
\multicolumn{6}{p{16.5cm}}{\scriptsize{\textbf{Note:} This table presents the estimated effect of exposure to mandatory high school courses on the 1991 Constitution on identification with Catholic, non-Catholic Christian, non-Christian, and secular affiliations, separating individuals located in the Central and Eastern regions and Bogota (Panel A) from those located in the Pacific and Atlantic regions (Panel B). All columns report estimates of Equation (\ref{baselineDiD}) using an age window of $[-3, 3]$. The regression sample is drawn from the 2015, 2017, 2019, and 2021 waves of the Political Culture Survey conducted by DANE. All specifications include cohort, high school, and survey fixed effects, as well as dummy variables for female and ethnic-minority status. Robust standard errors, clustered at the region level, are reported in parentheses, and wild-bootstrap confidence intervals are shown in square brackets.  *, **, and *** indicate statistical significance at the 10\%, 5\%, and 1\% levels, respectively.} }
\end{tabular}
}
\end{center}
\end{table}


\begin{figure}[H]
\begin{center}
\caption{Partial Correlation: Ethnic Minority Share and Location in Central Regions}
\label{fig_distributionreligident6tehnicminor}
\vspace{-0.3cm}
\includegraphics[width=14cm,height=8cm]{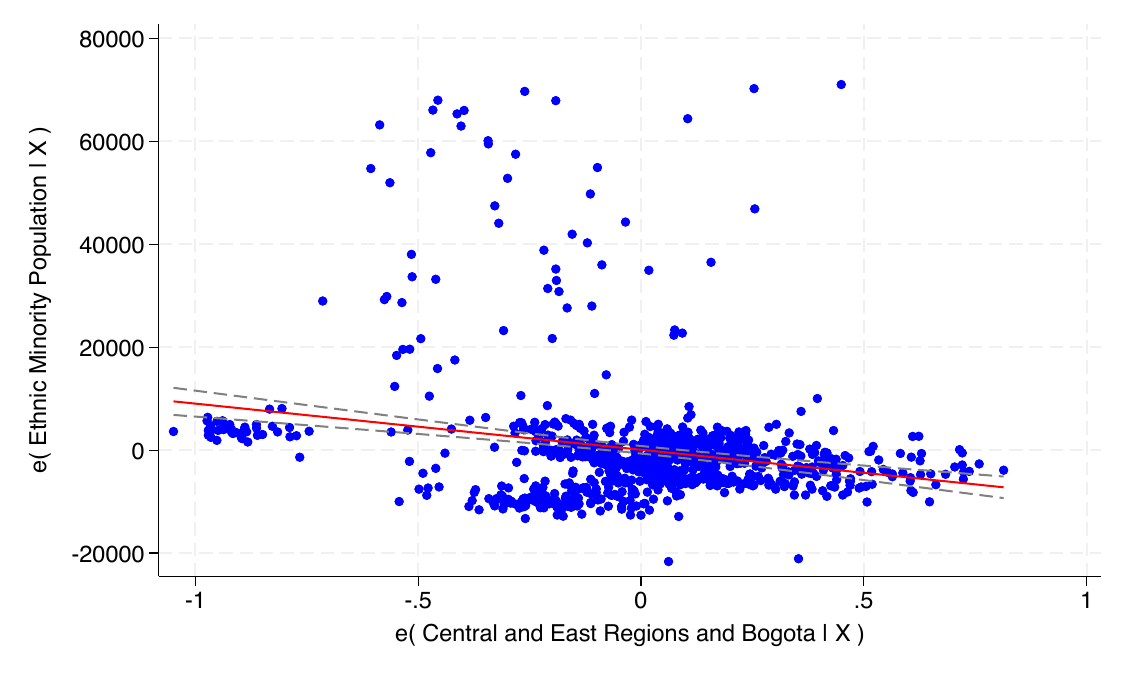}
 \begin{minipage}{14cm} \scriptsize\textbf{Note}: This figure presents a partial correlation plot illustrating the relationship between the share of the ethnic minority population per 100,000 inhabitants in 1993 and an indicator for whether the municipality belongs to the Central Regions (Central and Eastern regions and Bogota). The partial correlation is obtained from a regression that controls for the natural logarithm of total population (1995), municipal area, altitude, distance to the departmental capital, distance to Bogota, the share of rural population (1995), the Living Conditions Index (1995), and total municipal expenditure per capita (1995). Observations in the bottom and top 1\% of the dependent variable are excluded. Coef. = –8961.9 (SE = 4445.3). 
\end{minipage}
\end{center}
\end{figure}

\begin{table}[H]
\vspace{3cm}
\begin{center}
{
\renewcommand{\arraystretch}{0.8}
\setlength{\tabcolsep}{3pt}
\caption {Effect of Exposure to 1991 Constitution Courses on Religious Identification: Heterogeneous Effects by Ethnic Minority and Region}  \label{tab_effectallethnicityregion}
\vspace{-0.3cm}
\scriptsize
 \begin{tabular}{lccccc}
\hline\hline \addlinespace[0.15cm]
    & (1)& (2)& (3)  & (4)& (5)\\\addlinespace[0.12cm]\cmidrule[0.2pt](l){2-6}\addlinespace[0.12cm]
& \multicolumn{5}{c}{Dependent Variable: Identification with the following groups}\\\addlinespace[0.15cm]\cmidrule[0.2pt](l){2-6} \addlinespace[0.10cm]   

&\multicolumn{1}{c}{Catholic}&\multicolumn{1}{c}{Non-Catholic}&\multicolumn{1}{c}{Non-Christian}&\multicolumn{1}{c}{No Religion}&\multicolumn{1}{c}{Religious}\\

&\multicolumn{1}{c}{}&\multicolumn{1}{c}{Christian}&\multicolumn{1}{c}{}&\multicolumn{1}{c}{or Atheist}&\multicolumn{1}{c}{attendance}\\

\addlinespace[0.15cm]\hline\addlinespace[0.15cm]

\multicolumn{1}{l}{\emph{\underline{Panel A}:  }}      & \multicolumn{5}{c}{Ethnic Minority}\\\addlinespace[0.1cm]\cmidrule[0.2pt](l){2-6}\addlinespace[0.1cm]  
(Age $\leq$ 16 in 1994) $\times$ (High school) & -0.036 & 0.009 & 0.016 & 0.011 & 0.179 \\
  & (0.025) & (0.026) & (0.008) & (0.015) & (0.008)*** \\
  & [-0.121, 0.087] & [-0.023, 0.058] & [-0.060, 0.058] & [-0.115, 0.127] & [0.153, 0.310]*** \\
R-sq  & 0.030 & 0.025 & 0.021 & 0.038 & 0.102 \\
Observations  & 3,268 & 3,268 & 3,268 & 3,268 & 1,051 \\

\addlinespace[0.18cm]\hline\addlinespace[0.12cm]
\multicolumn{1}{l}{\emph{\underline{Panel B}:  }}      & \multicolumn{5}{c}{No Ethnic Minority}\\\addlinespace[0.1cm]\cmidrule[0.2pt](l){2-6}\addlinespace[0.1cm]  

(Age $\leq$ 16 in 1994) $\times$ (High school) & -0.031 & 0.020 & -0.005 & 0.016 & -0.060 \\
  & (0.008)** & (0.013) & (0.003) & (0.007)* & (0.024)* \\
  & [-0.045, -0.005]*** & [-0.024, 0.044] & [-0.015, 0.001]*** & [-0.009, 0.040] & [-0.140, -0.016]*** \\
R-sq  & 0.017 & 0.015 & 0.006 & 0.013 & 0.050 \\
Observations  & 16,607 & 16,607 & 16,607 & 16,607 & 7,158 \\

\addlinespace[0.18cm]\hline\addlinespace[0.12cm]
\multicolumn{1}{l}{\emph{\underline{Panel C}:  }}      & \multicolumn{5}{c}{Ethnic Minority and Pacific and Atlantic Regions}\\\addlinespace[0.1cm]\cmidrule[0.2pt](l){2-6}\addlinespace[0.1cm]  

(Age $\leq$ 16 in 1994) $\times$ (High school) & -0.024 & -0.001 & 0.024 & 0.001 & 0.201 \\
  & (0.028) & (0.032) & (0.004)* & (0.008) & (0.005)** \\
  & [-0.068, -0.005]*** & [-0.025, 0.051] & [0.020, 0.028]*** & [-0.009, 0.009] & [0.199, 0.203]*** \\
R-sq  & 0.007 & 0.004 & 0.007 & 0.013 & 0.047 \\
Observations  & 2,719 & 2,719 & 2,719 & 2,719 & 818 \\

\addlinespace[0.18cm]\hline\addlinespace[0.12cm]
\multicolumn{1}{l}{\emph{\underline{Panel D}:  }}      & \multicolumn{5}{c}{No Ethnic Minority and Pacific and Atlantic Regions}\\\addlinespace[0.1cm]\cmidrule[0.2pt](l){2-6}\addlinespace[0.1cm]  

(Age $\leq$ 16 in 1994) $\times$ (High school) & -0.043 & 0.047 & -0.008 & 0.004 & -0.034 \\
  & (0.007)* & (0.007)* & (0.007) & (0.007) & (0.010) \\
  & [-0.047, -0.031]*** & [0.038, 0.053]*** & [-0.016, -0.001]*** & [-0.006, 0.010] & [-0.045, -0.024]*** \\
R-sq  & 0.008 & 0.006 & 0.002 & 0.007 & 0.062 \\
Observations  & 6,614 & 6,614 & 6,614 & 6,614 & 2,668 \\

\addlinespace[0.18cm]\hline\addlinespace[0.12cm]
\multicolumn{1}{l}{\emph{\underline{Panel E}:  }}      & \multicolumn{5}{c}{Ethnic Minority and Central and East Regions and Bogota}\\\addlinespace[0.1cm]\cmidrule[0.2pt](l){2-6}\addlinespace[0.1cm]  

(Age $\leq$ 16 in 1994) $\times$ (High school) & -0.085 & 0.045 & -0.031 & 0.071 & 0.129 \\
  & (0.036) & (0.012)* & (0.015) & (0.056) & (0.047) \\
  & [-0.116, 0.020] & [0.022, 0.078]*** & [-0.048, 0.031] & [-0.090, 0.126] & [0.076, 0.245]*** \\
R-sq  & 0.040 & 0.032 & 0.031 & 0.055 & 0.027 \\
Observations  & 549 & 549 & 549 & 549 & 233 \\

\addlinespace[0.18cm]\hline\addlinespace[0.12cm]
\multicolumn{1}{l}{\emph{\underline{Panel F}:  }}      & \multicolumn{5}{c}{No Ethnic Minority and Central and East Regions and Bogota}\\\addlinespace[0.1cm]\cmidrule[0.2pt](l){2-6}\addlinespace[0.1cm]  

(Age $\leq$ 16 in 1994) $\times$ (High school) & -0.021 & -0.000 & -0.003 & 0.024 & -0.089 \\
  & (0.014) & (0.017) & (0.002) & (0.004)** & (0.042) \\
  & [-0.044, -0.004]*** & [-0.023, 0.025] & [-0.007, -0.000]*** & [0.019, 0.031]*** & [-0.158, -0.032]*** \\
R-sq  & 0.002 & 0.004 & 0.002 & 0.004 & 0.019 \\
Observations  & 9,993 & 9,993 & 9,993 & 9,993 & 4,490 \\

\addlinespace[0.15cm]\hline\hline\addlinespace[0.15cm]
\multicolumn{6}{p{16.5cm}}{\scriptsize{\textbf{Note:} This table reports the estimated effect of exposure to mandatory high school courses on the 1991 Constitution on Catholic, non-Catholic Christian, non-Christian, and secular identification, disaggregated by ethnic-minority status and region. All columns estimate Equation (\ref{baselineDiD}) using an age window of $[-3,3]$ and data from the 2015, 2017, 2019, and 2021 waves of DANE’s Political Culture Survey. Panel A shows results for ethnic-minority individuals (primarily Afro-Colombian or Indigenous); Panel B for non-minority individuals; Panels C and D for minority and non-minority individuals in the Pacific/Atlantic regions; and Panels E and F for minority and non-minority individuals in the Central/Eastern regions and Bogota. All specifications include cohort, high school, and survey fixed effects, as well as controls for gender. Robust standard errors clustered at the region level are shown in parentheses; wild-bootstrap confidence intervals appear in square brackets. *, **, and *** indicate statistical significance at the 10\%, 5\%, and 1\% levels, respectively.} }
\end{tabular}
}
\end{center}
\end{table}


\begin{table}[H]
\begin{center}
{
\renewcommand{\arraystretch}{0.8}
\setlength{\tabcolsep}{15pt}
\caption {Effect of Exposure to 1991 Constitution Courses on Religious Attendance, Political Ideology and Labor Market Participation}  \label{table_altermechanisms}
\vspace{-0.3cm}
\footnotesize
\centering  \begin{tabular}{lccc}
\hline\hline \addlinespace[0.15cm]
    & (1)& (2)& (3)\\\addlinespace[0.12cm]\hline\addlinespace[0.12cm]
      \multicolumn{1}{l}{\emph{\underline{Panel A}:  }}      & \multicolumn{3}{c}{Dep. Var: Religious Attendance}\\\addlinespace[0.1cm]\cmidrule[0.2pt](l){2-4}\addlinespace[0.1cm]               
(Age $\leq$ 16 in 1994) $\times$ (High school) & -0.031 & -0.028 & -0.025 \\
  & (0.036) & (0.034) & (0.032) \\
  & [-0.149, 0.035] & [-0.142, 0.035] & [-0.130, 0.034] \\
R-sq  & 0.003 & 0.007 & 0.048 \\
Observations  & 8,209 & 8,209 & 8,209 \\

\addlinespace[0.15cm]\hline\addlinespace[0.15cm]

      \multicolumn{1}{l}{\emph{\underline{Panel B}:  }}      & \multicolumn{3}{c}{Dep. Var: Right-Wing Ideology}\\\addlinespace[0.1cm]\cmidrule[0.2pt](l){2-4}\addlinespace[0.1cm]               
(Age $\leq$ 16 in 1994) $\times$ (High school) & -0.039 & -0.043 & -0.045 \\
  & (0.049) & (0.040) & (0.046) \\
  & [-0.201, 0.087] & [-0.176, 0.061] & [-0.141, 0.074] \\
R-sq  & 0.002 & 0.027 & 0.037 \\
Observations  & 16,029 & 16,029 & 16,029 \\

\addlinespace[0.15cm]\hline\addlinespace[0.15cm]

      \multicolumn{1}{l}{\emph{\underline{Panel C}:  }}      & \multicolumn{3}{c}{Dep. Var: Labor Market Participation}\\\addlinespace[0.1cm]\cmidrule[0.2pt](l){2-4}\addlinespace[0.1cm]               
(Age $\leq$ 16 in 1994) $\times$ (High school) & 0.026 & 0.024 & 0.026 \\
  & (0.020) & (0.020) & (0.016) \\
  & [-0.041, 0.066] & [-0.042, 0.066] & [-0.039, 0.056] \\
R-sq  & 0.020 & 0.032 & 0.198 \\
Observations  & 16,527 & 16,527 & 16,527 \\

    \addlinespace[0.15cm]\hline    \addlinespace[0.15cm]
\multicolumn{1}{l}{Cohort, region \& year fixed effects}     &  N& Y& Y \\
\multicolumn{1}{l}{Interacted fixed effects}    &  N& N& Y \\
\addlinespace[0.15cm]\hline\hline\addlinespace[0.15cm]
\multicolumn{4}{p{14.3cm}}{\scriptsize{
\textbf{Note:} This table reports the estimated effect of exposuThis table reports the estimated effect of exposure to mandatory high school courses on the 1991 Constitution on religious attendance, political ideology, and labor market participation. All columns report estimates from Equation~(\ref{baselineDiD}) using an age window of $[-3,3]$. The regression samples are drawn from the 2015, 2017, 2019, and 2021 waves of the Political Culture Survey conducted by DANE. The dependent variable in Panel A measures religious attendance, with higher values indicating more frequent participation in religious activities. The question asks: “How often do you attend meetings of the following voluntary organizations: churches, organizations, or religious groups? 1. Once a week; 2. Once or twice a month; 3. Once or twice a year; 4. Never.” A binary variable equals 1 if the respondent selects 1 or 2, and 0 otherwise. The dependent variable in Panel B measures political ideology on a 1–10 scale, where higher values indicate a more right-wing position. The question asks: “On a scale from 1 to 10, where 1 means left and 10 means right, where would you place yourself?” The dependent variable in Panel C captures labor market participation. The question asks: “What activity occupied most of your time last week?” A binary variable equals 1 if the respondent reported either working or looking for work.
Specifications in column (2) include fixed effects for age cohort, region, and survey year. 
Column (3) additionally includes dummy variables for female and ethnic-minority status, as well as cohort-by-region, high school-by-region, and survey-year-by-region fixed effects.  Robust standard errors clustered at the region level are reported in parentheses, and wild-bootstrap confidence intervals are shown in square brackets. 
*, **, and *** indicate statistical significance at the 10\%, 5\%, and 1\% levels, respectively.
}}

\end{tabular}
}
\end{center}
\end{table}

\newpage
\hbox {} 
\appendix
\section{Appendix}
\setcounter{table}{0}
\setcounter{figure}{0}
\renewcommand{\thefigure}{\Alph{section}\arabic{figure}}
\renewcommand{\thetable}{\Alph{section}\arabic{table}}

\subsection{Additional Figures and Tables}

\begin{table}[H]
\vspace{1cm}
\begin{center}
{
\renewcommand{\arraystretch}{0.8}
\setlength{\tabcolsep}{8pt}
\caption {Effect of Exposure to 1991 Constitution Courses on Religious Identification
After 1994}  \label{tab_effectallpost1994}
\vspace{-0.3cm}
\scriptsize
 \begin{tabular}{lccccccccc}
\hline\hline \addlinespace[0.15cm]
    & (1)& (2)& (3)  & (4)\\\addlinespace[0.12cm]\cmidrule[0.2pt](l){2-5}\addlinespace[0.12cm]
& \multicolumn{4}{c}{Dependent Variable: Identification with }\\\addlinespace[0.15cm]\cmidrule[0.2pt](l){2-5} \addlinespace[0.10cm]   

&\multicolumn{1}{c}{Catholic}&\multicolumn{1}{c}{Non-Catholic}&\multicolumn{1}{c}{Non-Christian}&\multicolumn{1}{c}{No Religion}\\

&\multicolumn{1}{c}{}&\multicolumn{1}{c}{Christian}&\multicolumn{1}{c}{}&\multicolumn{1}{c}{or Atheist}\\

\addlinespace[0.1cm]\cmidrule[0.2pt](l){2-2} \cmidrule[0.2pt](l){3-3}\cmidrule[0.2pt](l){4-4}\cmidrule[0.2pt](l){5-5}\addlinespace[0.12cm]

(Age $\leq$ 16 in 1996) $\times$ (High school) & -0.005 & -0.001 & 0.000 & 0.005 \\
  & (0.015) & (0.011) & (0.004) & (0.005) \\
  & [-0.055, 0.023] & [-0.037, 0.036] & [-0.008, 0.032] & [-0.009, 0.044] \\
R-sq  & 0.020 & 0.017 & 0.005 & 0.015 \\
Observations  & 20,653 & 20,653 & 20,653 & 20,653 \\
\addlinespace[0.1cm]\cmidrule[0.2pt](l){2-2} \cmidrule[0.2pt](l){3-3}\cmidrule[0.2pt](l){4-4}\cmidrule[0.2pt](l){5-5}\addlinespace[0.10cm]    

(Age $\leq$ 16 in 1998) $\times$ (High school) & 0.032 & -0.007 & 0.007 & -0.032 \\
  & (0.018) & (0.014) & (0.004) & (0.010)** \\
  & [-0.007, 0.084] & [-0.097, 0.030] & [-0.006, 0.023] & [-0.063, 0.035]* \\
R-sq  & 0.022 & 0.018 & 0.005 & 0.018 \\
Observations  & 21,406 & 21,406 & 21,406 & 21,406 \\
\addlinespace[0.1cm]\cmidrule[0.2pt](l){2-2} \cmidrule[0.2pt](l){3-3}\cmidrule[0.2pt](l){4-4}\cmidrule[0.2pt](l){5-5}\addlinespace[0.10cm]

(Age $\leq$ 16 in 2000) $\times$ (High school) & 0.011 & 0.002 & 0.000 & -0.013 \\
  & (0.007) & (0.014) & (0.004) & (0.012) \\
  & [-0.013, 0.082] & [-0.056, 0.049] & [-0.013, 0.008] & [-0.047, 0.011] \\
R-sq  & 0.022 & 0.016 & 0.005 & 0.019 \\
Observations  & 20,935 & 20,935 & 20,935 & 20,935 \\
\addlinespace[0.1cm]\cmidrule[0.2pt](l){2-2} \cmidrule[0.2pt](l){3-3}\cmidrule[0.2pt](l){4-4}\cmidrule[0.2pt](l){5-5}\addlinespace[0.10cm]

(Age $\leq$ 16 in 2002) $\times$ (High school) & -0.022 & -0.001 & -0.003 & 0.026 \\
  & (0.012) & (0.014) & (0.002) & (0.005)*** \\
  & [-0.061, 0.018] & [-0.046, 0.064] & [-0.009, 0.005] & [-0.005, 0.039]* \\
R-sq  & 0.017 & 0.014 & 0.005 & 0.018 \\
Observations  & 20,912 & 20,912 & 20,912 & 20,912 \\
\addlinespace[0.1cm]\cmidrule[0.2pt](l){2-2} \cmidrule[0.2pt](l){3-3}\cmidrule[0.2pt](l){4-4}\cmidrule[0.2pt](l){5-5}\addlinespace[0.10cm]

(Age $\leq$ 16 in 2004) $\times$ (High school) & -0.045 & 0.028 & 0.003 & 0.014 \\
  & (0.010)** & (0.010)* & (0.003) & (0.005)** \\
  & [-0.071, -0.011]*** & [-0.007, 0.047] & [-0.007, 0.009] & [0.007, 0.045]*** \\
R-sq  & 0.020 & 0.016 & 0.005 & 0.017 \\
Observations  & 21,400 & 21,400 & 21,400 & 21,400 \\

\addlinespace[0.15cm]\hline\hline\addlinespace[0.15cm]
\multicolumn{5}{p{15cm}}{\scriptsize{\textbf{Note:} This table presents the estimated effect of exposure to mandatory high school courses on the 1991 Constitution on identification with Catholic, non-Catholic Christian, non-Christian, and secular affiliations. All columns report estimates from Equation (\ref{baselineDiD}), but instead of using the indicator for being at most 16 years old in 1994, we replace it with an indicator for being that age in the specific year indicated in each row. All columns report estimates of Equation (\ref{baselineDiD}) using an age window of $[-3, 3]$. The regression sample is drawn from the 2015, 2017, 2019, and 2021 waves of the Political Culture Survey conducted by DANE. All specifications include cohort-by-region, high school-by-region, and survey-year-by-region fixed effects, as well as dummy variables for female and ethnic-minority status. Robust standard errors, clustered at the region level, are reported in parentheses, and wild-bootstrap confidence intervals are shown in square brackets.  *, **, and *** indicate statistical significance at the 10\%, 5\%, and 1\% levels, respectively.} }
\end{tabular}
}
\end{center}
\end{table}


\begin{table}[H]
\begin{center}
\renewcommand{\arraystretch}{0.9}
\setlength{\tabcolsep}{10pt}
\caption {Protestant missionary agencies entering Colombia, 1886–1930}  \label{tab_missions_1886_1930}
\vspace{-0.3cm}
\begin{tabular}{lcc}
\addlinespace[0.15cm]\hline\hline\addlinespace[0.15cm]
Denomination/Agency & Year & Initial location
\\\addlinespace[0.15cm]\cmidrule[0.2pt](l){1-1} \cmidrule[0.2pt](l){2-2} \cmidrule[0.2pt](l){3-3} 
Unión Misionera Evangélica & 1908 & Cali y Palmira \\
Sociedad Bíblica Americana & 1912 & Cartagena \\
Sociedad Bíblica Británica y Extranjera & 1917 & Pasto \\
Iglesia Adventista del Séptimo Día & 1921 & Bogota, Medellín, Barranquilla, Cali \\
Alianza Misionera Escandinava & 1922 & Cúcuta \\
Alianza Cristiana y Misionera & 1925 & Ipiales \\
Instituto Colombo–Venezolano & 1926 & Unknown \\
Misión Presbiteriana Cumberland & 1927 & Cali \\
Convención Bautista Nacional & 1929 & Santa Marta y Ciénaga \\
\addlinespace[0.15cm]\hline\hline\addlinespace[0.15cm]
\end{tabular}
 \begin{minipage}{15cm} \footnotesize\textbf{Note}:  Reproduced from  \cite{Beltran2013} [Table~1, p.53].
\end{minipage}
\end{center}
\end{table}

\begin{table}[H]
\begin{center}
{
\renewcommand{\arraystretch}{0.7}
\setlength{\tabcolsep}{10pt}
\caption {Correlation Between Location in Central Regions and Either Catholic Church Density or Ethnic Minority Population Share}  \label{tab_corrcathethniregions}
\vspace{-0.3cm}
\small
\centering  \begin{tabular}{lcccc}
\hline\hline \addlinespace[0.15cm]
    & (1)& (2)& (3)& (4)\\\addlinespace[0.1cm]\cmidrule[0.2pt](l){2-5}\addlinespace[0.10cm]
   & \multicolumn{4}{c}{Dependent Variable:}\\\addlinespace[0.1cm]\cmidrule[0.2pt](l){2-5}\addlinespace[0.1cm]               
   & \multicolumn{2}{c}{Catholic Churches}& \multicolumn{2}{c}{Ethnic Minority} \\
      & \multicolumn{2}{c}{per 100{,}000 }& \multicolumn{2}{c}{Population Share} \\\addlinespace[0.10cm]\cmidrule[0.2pt](l){2-3}\cmidrule[0.2pt](l){4-5}\addlinespace[0.1cm]               
Bogota–Central–Eastern (=1)&       8.396***&       3.440***&     -9848.4** &     -9513.4*  \\
                    &      (4.14)   &      (3.42)   &     (-2.15)   &     (-2.04)   \\
R-sq                &       0.085   &       0.417   &       0.104   &       0.223   \\
Observations        &         951   &         937   &         984   &         968   \\

\addlinespace[0.15cm]\hline\addlinespace[0.15cm]
\multicolumn{1}{l}{Controls}    &  N& Y & N& Y \\
\addlinespace[0.15cm]\hline\hline\addlinespace[0.15cm]
\multicolumn{5}{p{13.5cm}}{\scriptsize{\textbf{Note:} This table reports correlations between an indicator for whether a municipality belongs to the Central Regions (Central and Eastern regions and Bogota) versus the Peripheral regions (Atlantic and Pacific) and either Catholic Church density or the ethnic-minority population share. Columns (1)–(2) show correlations with the number of Catholic churches per 100,000 inhabitants in 1995; Columns (3)–(4) show correlations with the share of the ethnic-minority population per 100,000 inhabitants in 1993. Columns (2) and (4) include controls for the logarithm of total population (1995), municipal area, altitude, distance to the departmental capital, distance to Bogota, the share of rural population (1995), the Living Conditions Index (1995), and total municipal expenditure per capita (1995). Data on Catholic churches, the Living Conditions Index, and municipal expenditure come from Fundación Social. Ethnic-minority data come from the 1993 Population Census (DANE). Population data (total and rural) come from DANE, and municipal area, altitude, and distance measures come from CEDE. Robust standard errors, clustered at the department level, are reported in parentheses. *, **, and *** indicate statistical significance at the 10\%, 5\%, and 1\% levels, respectively.} }
\end{tabular}
}
\end{center}
\end{table}


\begin{table}[H]
\vspace{3cm}
\begin{center}
{
\renewcommand{\arraystretch}{0.8}
\setlength{\tabcolsep}{8pt}
\caption {Effect of Exposure to 1991 Constitution Courses on Religious Identification: Heterogeneity by Gender}  \label{tab_effectallgender}
\vspace{-0.3cm}
\scriptsize
 \begin{tabular}{lccccccccc}
\hline\hline \addlinespace[0.15cm]
    & (1)& (2)& (3)  & (4)\\\addlinespace[0.12cm]\cmidrule[0.2pt](l){2-5}\addlinespace[0.12cm]
& \multicolumn{4}{c}{Dependent Variable: Identification with the following groups}\\\addlinespace[0.15cm]\cmidrule[0.2pt](l){2-5} \addlinespace[0.10cm]   

&\multicolumn{1}{c}{Catholic}&\multicolumn{1}{c}{Non-Catholic}&\multicolumn{1}{c}{Non-Christian}&\multicolumn{1}{c}{No Religion}\\

&\multicolumn{1}{c}{}&\multicolumn{1}{c}{Christian}&\multicolumn{1}{c}{}&\multicolumn{1}{c}{or Atheist}\\

\addlinespace[0.15cm]\hline\addlinespace[0.15cm]

\multicolumn{1}{l}{\emph{\underline{Panel A}:  }}      & \multicolumn{4}{c}{Women}\\\addlinespace[0.1cm]\cmidrule[0.2pt](l){2-5}\addlinespace[0.1cm]  
(Age $\leq$ 16 in 1994) $\times$ (High school) & -0.023 & 0.017 & -0.005 & 0.011 \\
  & (0.014) & (0.016) & (0.003) & (0.010) \\
  & [-0.090, 0.003] & [-0.025, 0.083] & [-0.015, 0.011] & [-0.021, 0.036] \\
R-sq  & 0.025 & 0.023 & 0.008 & 0.012 \\
Observations  & 10,728 & 10,728 & 10,728 & 10,728 \\

\addlinespace[0.18cm]\hline\addlinespace[0.12cm]
\multicolumn{1}{l}{\emph{\underline{Panel B}:  }}      & \multicolumn{4}{c}{Men}\\\addlinespace[0.1cm]\cmidrule[0.2pt](l){2-5}\addlinespace[0.1cm]  

(Age $\leq$ 16 in 1994) $\times$ (High school) & -0.046 & 0.021 & 0.004 & 0.021 \\
  & (0.016)** & (0.016) & (0.004) & (0.013) \\
  & [-0.099, 0.018]* & [-0.070, 0.076] & [-0.014, 0.015] & [-0.011, 0.063]* \\
R-sq  & 0.024 & 0.015 & 0.006 & 0.017 \\
Observations  & 9,147 & 9,147 & 9,147 & 9,147 \\

\addlinespace[0.15cm]\hline\hline\addlinespace[0.15cm]
\multicolumn{5}{p{14.5cm}}{\scriptsize{\textbf{Note:}  This table presents the estimated effect of exposure to mandatory high school courses on the 1991 Constitution on identification with Catholic, non-Catholic Christian, non-Christian, and secular affiliations, separating results for women (Panel A) and men (Panel B). All columns report estimates of Equation~(\ref{baselineDiD}) using an age window of $[-3, 3]$. The regression sample is drawn from the 2015, 2017, 2019, and 2021 waves of the Political Culture Survey conducted by DANE. All specifications include cohort-by-region, high school-by-region, and survey-year-by-region fixed effects, as well as a dummy variable for ethnic-minority status. Robust standard errors, clustered at the region level, are reported in parentheses, and wild-bootstrap confidence intervals are shown in square brackets. *, **, and *** indicate statistical significance at the 10\%, 5\%, and 1\% levels, respectively.} }
\end{tabular}
}
\end{center}
\end{table}

\begin{table}[H]
\begin{center}
{
\renewcommand{\arraystretch}{0.7}
\setlength{\tabcolsep}{10pt}
\caption {Effect on Catholic Identification (Years Prior to 1994 and Age Windows $[-2,2]$, $[-3,3]$, and $[-4,4]$)}  \label{tab_efftrcatholi_before94_w234}
\vspace{-0.3cm}
\small
\centering  \begin{tabular}{lccc}
\hline\hline \addlinespace[0.15cm]
    & (1)& (2)& (3)\\\addlinespace[0.1cm]\cmidrule[0.2pt](l){2-4}\addlinespace[0.10cm]

   & \multicolumn{3}{c}{Dep. Var: Catholic Identification}\\\addlinespace[0.1cm]\cmidrule[0.2pt](l){2-4}\addlinespace[0.1cm]               
   & \multicolumn{1}{c}{[-2,2]}& \multicolumn{1}{c}{[-3,3]}& \multicolumn{1}{c}{[-4,4]}\\\addlinespace[0.1cm]\cmidrule[0.2pt](l){1-1}\cmidrule[0.2pt](l){2-2}\cmidrule[0.2pt](l){3-3}\cmidrule[0.2pt](l){4-4}\addlinespace[0.1cm]               
(Age $\leq$ 16 in 1992) $\times$ (Highschool) & -0.021 & -0.016 & -0.017 \\
  & (0.017) & (0.009) & (0.008) \\
  & [-0.052, 0.047] & [-0.033, 0.016] & [-0.032, 0.010] \\
R-sq  & 0.020 & 0.018 & 0.018 \\
Observations  & 13,072 & 18,577 & 24,402 \\

\addlinespace[0.1cm]\cmidrule[0.2pt](l){1-1}\cmidrule[0.2pt](l){2-2}\cmidrule[0.2pt](l){3-3}\cmidrule[0.2pt](l){4-4}\addlinespace[0.1cm]      
(Age $\leq$ 16 in 1990) $\times$ (Highschool) & 0.021 & 0.022 & 0.020 \\
  & (0.012) & (0.013) & (0.011) \\
  & [-0.010, 0.087]* & [-0.012, 0.115] & [-0.001, 0.098]* \\
R-sq  & 0.019 & 0.019 & 0.018 \\
Observations  & 12,920 & 18,196 & 23,520 \\

\addlinespace[0.1cm]\cmidrule[0.2pt](l){1-1}\cmidrule[0.2pt](l){2-2}\cmidrule[0.2pt](l){3-3}\cmidrule[0.2pt](l){4-4}\addlinespace[0.1cm]      
(Age $\leq$ 16 in 1988) $\times$ (Highschool) & 0.003 & 0.014 & 0.009 \\
  & (0.014) & (0.011) & (0.011) \\
  & [-0.043, 0.099] & [-0.016, 0.086] & [-0.020, 0.094] \\\
R-sq  & 0.015 & 0.016 & 0.017 \\
Observations  & 12,921 & 18,350 & 23,516 \\

\addlinespace[0.1cm]\cmidrule[0.2pt](l){1-1}\cmidrule[0.2pt](l){2-2}\cmidrule[0.2pt](l){3-3}\cmidrule[0.2pt](l){4-4}\addlinespace[0.1cm]      
(Age $\leq$ 16 in 1986) $\times$ (Highschool) & -0.003 & -0.006 & -0.011 \\
  & (0.011) & (0.014) & (0.012) \\
  & [-0.023, 0.043] & [-0.062, 0.040] & [-0.044, 0.020] \\
R-sq  & 0.018 & 0.016 & 0.016 \\
Observations  & 13,135 & 18,245 & 23,279 \\

\addlinespace[0.1cm]\cmidrule[0.2pt](l){1-1}\cmidrule[0.2pt](l){2-2}\cmidrule[0.2pt](l){3-3}\cmidrule[0.2pt](l){4-4}\addlinespace[0.1cm]      
(Age $\leq$ 16 in 1984) $\times$ (Highschool) & -0.019 & -0.011 & -0.009 \\
  & (0.014) & (0.010) & (0.013) \\
  & [-0.096, 0.007] & [-0.067, 0.008] & [-0.090, 0.014] \\
R-sq  & 0.018 & 0.018 & 0.019 \\
Observations  & 13,084 & 18,472 & 23,535 \\

\addlinespace[0.15cm]\hline\hline\addlinespace[0.15cm]
\multicolumn{4}{p{14cm}}{\scriptsize{\textbf{Note:} All columns present estimates of Equation (\ref{baselineDiD}), but instead of using the indicator for being at most 16 years old in 1994, we replace it with an indicator for being that age in the specific year listed in each row. Column (1) uses an age window of $[-2,2]$, column (2) uses $[-3,3]$, and column (3) uses $[-4,4]$. The regression sample is drawn from the 2015, 2017, 2019, and 2021 waves of the Political Culture Survey conducted by DANE. All specifications include cohort-by-region, high school-by-region, and survey year-by-region fixed effects, as well as a dummy variable for ethnic-minority status. Robust standard errors, clustered at the region level, are reported in parentheses, and wild-bootstrap confidence intervals are shown in square brackets. *, **, and *** indicate statistical significance at the 10\%, 5\%, and 1\% levels, respectively.} }
\end{tabular}
}
\end{center}
\end{table}

\begin{table}[H]
\begin{center}
{
\renewcommand{\arraystretch}{0.7}
\setlength{\tabcolsep}{10pt}
\caption {Effect on Identification with Non-Catholic Christian Denominations (Years Prior to 1994 and Age Windows $[-2,2]$, $[-3,3]$, and $[-4,4]$)}  \label{tab_efftrnocatchriall_before94_w234}
\vspace{-0.3cm}
\small
\centering  \begin{tabular}{lccc}
\hline\hline \addlinespace[0.15cm]
    & (1)& (2)& (3)\\\addlinespace[0.1cm]\cmidrule[0.2pt](l){2-4}\addlinespace[0.10cm]

   & \multicolumn{3}{c}{Dep. Var: Non-Catholic Chrsitian Identification}\\\addlinespace[0.1cm]\cmidrule[0.2pt](l){2-4}\addlinespace[0.1cm]               
   & \multicolumn{1}{c}{[-2,2]}& \multicolumn{1}{c}{[-3,3]}& \multicolumn{1}{c}{[-4,4]}\\\addlinespace[0.1cm]\cmidrule[0.2pt](l){1-1}\cmidrule[0.2pt](l){2-2}\cmidrule[0.2pt](l){3-3}\cmidrule[0.2pt](l){4-4}\addlinespace[0.1cm]               
(Age $\leq$ 16 in 1992) $\times$ (Highschool) & 0.019 & 0.017 & 0.019 \\
  & (0.013) & (0.007)* & (0.010) \\
  & [-0.014, 0.062] & [0.006, 0.041]*** & [-0.005, 0.051] \\
R-sq  & 0.015 & 0.014 & 0.014 \\
Observations  & 13,072 & 18,577 & 24,402 \\

\addlinespace[0.1cm]\cmidrule[0.2pt](l){1-1}\cmidrule[0.2pt](l){2-2}\cmidrule[0.2pt](l){3-3}\cmidrule[0.2pt](l){4-4}\addlinespace[0.1cm]      
(Age $\leq$ 16 in 1990) $\times$ (Highschool) & -0.011 & -0.008 & -0.004 \\
  & (0.007) & (0.012) & (0.010) \\
  & [-0.077, 0.011] & [-0.066, 0.033] & [-0.051, 0.021] \\
R-sq  & 0.014 & 0.014 & 0.013 \\
Observations  & 12,920 & 18,196 & 23,520 \\

\addlinespace[0.1cm]\cmidrule[0.2pt](l){1-1}\cmidrule[0.2pt](l){2-2}\cmidrule[0.2pt](l){3-3}\cmidrule[0.2pt](l){4-4}\addlinespace[0.1cm]      
(Age $\leq$ 16 in 1988) $\times$ (Highschool) & 0.002 & 0.000 & 0.007 \\
  & (0.007) & (0.004) & (0.006) \\
  & [-0.021, 0.014] & [-0.015, 0.013] & [-0.017, 0.021] \\
R-sq  & 0.010 & 0.011 & 0.012 \\
Observations  & 12,921 & 18,350 & 23,516 \\

\addlinespace[0.1cm]\cmidrule[0.2pt](l){1-1}\cmidrule[0.2pt](l){2-2}\cmidrule[0.2pt](l){3-3}\cmidrule[0.2pt](l){4-4}\addlinespace[0.1cm]      
(Age $\leq$ 16 in 1986) $\times$ (Highschool) & 0.013 & 0.013 & 0.012 \\
  & (0.008) & (0.012) & (0.012) \\
  & [-0.008, 0.046] & [-0.008, 0.060] & [-0.010, 0.053] \\
R-sq  & 0.012 & 0.012 & 0.011 \\
Observations  & 13,135 & 18,245 & 23,279 \\

\addlinespace[0.1cm]\cmidrule[0.2pt](l){1-1}\cmidrule[0.2pt](l){2-2}\cmidrule[0.2pt](l){3-3}\cmidrule[0.2pt](l){4-4}\addlinespace[0.1cm]      
(Age $\leq$ 16 in 1984) $\times$ (Highschool) & 0.005 & -0.001 & 0.002 \\
  & (0.010) & (0.009) & (0.010) \\
  & [-0.022, 0.033] & [-0.023, 0.020] & [-0.019, 0.049] \\
R-sq  & 0.013 & 0.013 & 0.013 \\
Observations  & 13,084 & 18,472 & 23,535 \\

\addlinespace[0.15cm]\hline\hline\addlinespace[0.15cm]
\multicolumn{4}{p{14cm}}{\scriptsize{\textbf{Note:}  All columns present estimates of Equation (\ref{baselineDiD}), but instead of using the indicator for being at most 16 years old in 1994, we replace it with an indicator for being that age in the specific year listed in each row. Column (1) uses an age window of $[-2,2]$, column (2) uses $[-3,3]$, and column (3) uses $[-4,4]$. The regression sample is drawn from the 2015, 2017, 2019, and 2021 waves of the Political Culture Survey conducted by DANE. All specifications include cohort-by-region, high school-by-region, and survey year-by-region fixed effects, as well as a dummy variable for ethnic-minority status. Robust standard errors, clustered at the region level, are reported in parentheses, and wild-bootstrap confidence intervals are shown in square brackets. *, **, and *** indicate statistical significance at the 10\%, 5\%, and 1\% levels, respectively.} }
\end{tabular}
}
\end{center}
\end{table}

\begin{table}[H]
\begin{center}
{
\renewcommand{\arraystretch}{0.7}
\setlength{\tabcolsep}{10pt}
\caption {Effect on Identification with Non-Christian Religions (Years Prior to 1994 and Age Windows $[-2,2]$, $[-3,3]$, and $[-4,4]$)}  \label{tab_efftrnochristian_before94_w234}
\vspace{-0.3cm}
\small
\centering  \begin{tabular}{lccc}
\hline\hline \addlinespace[0.15cm]
    & (1)& (2)& (3)\\\addlinespace[0.1cm]\cmidrule[0.2pt](l){2-4}\addlinespace[0.10cm]

   & \multicolumn{3}{c}{Dep. Var: non-Christian Identification}\\\addlinespace[0.1cm]\cmidrule[0.2pt](l){2-4}\addlinespace[0.1cm]               
   & \multicolumn{1}{c}{[-2,2]}& \multicolumn{1}{c}{[-3,3]}& \multicolumn{1}{c}{[-4,4]}\\\addlinespace[0.1cm]\cmidrule[0.2pt](l){1-1}\cmidrule[0.2pt](l){2-2}\cmidrule[0.2pt](l){3-3}\cmidrule[0.2pt](l){4-4}\addlinespace[0.1cm]               
(Age $\leq$ 16 in 1992) $\times$ (Highschool) & 0.003 & -0.001 & -0.002 \\
  & (0.004) & (0.002) & (0.001)* \\
  & [-0.009, 0.025] & [-0.009, 0.003] & [-0.005, 0.000]*** \\
R-sq  & 0.006 & 0.005 & 0.005 \\
Observations  & 13,072 & 18,577 & 24,402 \\

\addlinespace[0.1cm]\cmidrule[0.2pt](l){1-1}\cmidrule[0.2pt](l){2-2}\cmidrule[0.2pt](l){3-3}\cmidrule[0.2pt](l){4-4}\addlinespace[0.1cm]      
(Age $\leq$ 16 in 1990) $\times$ (Highschool) & -0.005 & -0.005 & -0.003 \\
  & (0.004) & (0.004) & (0.004) \\
  & [-0.035, 0.002] & [-0.030, 0.005] & [-0.030, 0.006] \\
R-sq  & 0.006 & 0.005 & 0.004 \\
Observations  & 12,920 & 18,196 & 23,520 \\

\addlinespace[0.1cm]\cmidrule[0.2pt](l){1-1}\cmidrule[0.2pt](l){2-2}\cmidrule[0.2pt](l){3-3}\cmidrule[0.2pt](l){4-4}\addlinespace[0.1cm]      
(Age $\leq$ 16 in 1988) $\times$ (Highschool) & -0.000 & -0.002 & -0.002 \\
  & (0.006) & (0.006) & (0.005) \\
  & [-0.024, 0.015] & [-0.025, 0.014] & [-0.023, 0.012] \\
R-sq  & 0.004 & 0.004 & 0.004 \\
Observations  & 12,921 & 18,350 & 23,516 \\

\addlinespace[0.1cm]\cmidrule[0.2pt](l){1-1}\cmidrule[0.2pt](l){2-2}\cmidrule[0.2pt](l){3-3}\cmidrule[0.2pt](l){4-4}\addlinespace[0.1cm]      
(Age $\leq$ 16 in 1986) $\times$ (Highschool) & 0.001 & 0.004 & 0.003 \\
  & (0.003) & (0.002) & (0.002) \\
  & [-0.005, 0.009] & [-0.002, 0.007] & [-0.004, 0.008] \\
R-sq  & 0.005 & 0.004 & 0.004 \\
Observations  & 13,135 & 18,245 & 23,279 \\

\addlinespace[0.1cm]\cmidrule[0.2pt](l){1-1}\cmidrule[0.2pt](l){2-2}\cmidrule[0.2pt](l){3-3}\cmidrule[0.2pt](l){4-4}\addlinespace[0.1cm]      
(Age $\leq$ 16 in 1984) $\times$ (Highschool) & 0.005 & 0.005 & 0.004 \\
  & (0.004) & (0.003) & (0.003) \\
  & [-0.008, 0.025] & [-0.004, 0.015] & [-0.006, 0.017] \\
R-sq  & 0.005 & 0.004 & 0.004 \\
Observations  & 13,084 & 18,472 & 23,535 \\

\addlinespace[0.15cm]\hline\hline\addlinespace[0.15cm]
\multicolumn{4}{p{14cm}}{\scriptsize{\textbf{Note:} All columns present estimates of Equation (\ref{baselineDiD}), but instead of using the indicator for being at most 16 years old in 1994, we replace it with an indicator for being that age in the specific year listed in each row. Column (1) uses an age window of $[-2,2]$, column (2) uses $[-3,3]$, and column (3) uses $[-4,4]$. The regression sample is drawn from the 2015, 2017, 2019, and 2021 waves of the Political Culture Survey conducted by DANE. All specifications include cohort-by-region, high school-by-region, and survey year-by-region fixed effects, as well as a dummy variable for ethnic-minority status. Robust standard errors, clustered at the region level, are reported in parentheses, and wild-bootstrap confidence intervals are shown in square brackets.  *, **, and *** indicate statistical significance at the 10\%, 5\%, and 1\% levels, respectively.} }
\end{tabular}
}
\end{center}
\end{table}

\begin{table}[H]
\begin{center}
{
\renewcommand{\arraystretch}{0.7}
\setlength{\tabcolsep}{10pt}
\caption {Effect  on  Identification with No Religion or Atheism (Years Prior to 1994 and Age Windows $[-2,2]$, $[-3,3]$, and $[-4,4]$)}  \label{tab_efftrnoreligion_before94_w234}
\vspace{-0.3cm}
\small
\centering  \begin{tabular}{lccc}
\hline\hline \addlinespace[0.15cm]
    & (1)& (2)& (3)\\\addlinespace[0.1cm]\cmidrule[0.2pt](l){2-4}\addlinespace[0.10cm]

   & \multicolumn{3}{c}{Dep. Var: non-Religion or Atheist}\\\addlinespace[0.1cm]\cmidrule[0.2pt](l){2-4}\addlinespace[0.1cm]               
   & \multicolumn{1}{c}{[-2,2]}& \multicolumn{1}{c}{[-3,3]}& \multicolumn{1}{c}{[-4,4]}\\\addlinespace[0.1cm]\cmidrule[0.2pt](l){1-1}\cmidrule[0.2pt](l){2-2}\cmidrule[0.2pt](l){3-3}\cmidrule[0.2pt](l){4-4}\addlinespace[0.1cm]               
(Age $\leq$ 16 in 1992) $\times$ (Highschool) & -0.001 & -0.001 & 0.000 \\
  & (0.011) & (0.007) & (0.006) \\
  & [-0.023, 0.022] & [-0.013, 0.019] & [-0.014, 0.016] \\
R-sq  & 0.014 & 0.014 & 0.013 \\
Observations  & 13,072 & 18,577 & 24,402 \\

\addlinespace[0.1cm]\cmidrule[0.2pt](l){1-1}\cmidrule[0.2pt](l){2-2}\cmidrule[0.2pt](l){3-3}\cmidrule[0.2pt](l){4-4}\addlinespace[0.1cm]      
(Age $\leq$ 16 in 1990) $\times$ (Highschool) & -0.006 & -0.009 & -0.012 \\
  & (0.008) & (0.005) & (0.004)** \\
  & [-0.032, 0.014] & [-0.021, 0.002]* & [-0.021, -0.004]*** \\
R-sq  & 0.014 & 0.015 & 0.015 \\
Observations  & 12,920 & 18,196 & 23,520 \\

\addlinespace[0.1cm]\cmidrule[0.2pt](l){1-1}\cmidrule[0.2pt](l){2-2}\cmidrule[0.2pt](l){3-3}\cmidrule[0.2pt](l){4-4}\addlinespace[0.1cm]      
(Age $\leq$ 16 in 1988) $\times$ (Highschool) & -0.004 & -0.012 & -0.014 \\
  & (0.008) & (0.007) & (0.007) \\
  & [-0.062, 0.017] & [-0.051, 0.011]* & [-0.055, 0.008]* \\
R-sq  & 0.015 & 0.014 & 0.014 \\
Observations  & 12,921 & 18,350 & 23,516 \\

\addlinespace[0.1cm]\cmidrule[0.2pt](l){1-1}\cmidrule[0.2pt](l){2-2}\cmidrule[0.2pt](l){3-3}\cmidrule[0.2pt](l){4-4}\addlinespace[0.1cm]      
(Age $\leq$ 16 in 1986) $\times$ (Highschool) & -0.011 & -0.010 & -0.005 \\
  & (0.006) & (0.007) & (0.003) \\
  & [-0.030, -0.001]*** & [-0.039, -0.000]*** & [-0.013, 0.001]* \\
R-sq  & 0.015 & 0.013 & 0.014 \\
Observations  & 13,135 & 18,245 & 23,279 \\

\addlinespace[0.1cm]\cmidrule[0.2pt](l){1-1}\cmidrule[0.2pt](l){2-2}\cmidrule[0.2pt](l){3-3}\cmidrule[0.2pt](l){4-4}\addlinespace[0.1cm]      
(Age $\leq$ 16 in 1984) $\times$ (Highschool) & 0.010 & 0.007 & 0.004 \\
  & (0.006) & (0.006) & (0.004) \\
  & [-0.003, 0.047] & [-0.008, 0.047] & [-0.004, 0.026] \\
R-sq  & 0.015 & 0.016 & 0.016 \\
Observations  & 13,084 & 18,472 & 23,535 \\

\addlinespace[0.15cm]\hline\hline\addlinespace[0.15cm]
\multicolumn{4}{p{15.3cm}}{\scriptsize{\textbf{Note:}  All columns present estimates of Equation (\ref{baselineDiD}), but instead of using the indicator for being at most 16 years old in 1994, we replace it with an indicator for being that age in the specific year listed in each row. Column (1) uses an age window of $[-2,2]$, column (2) uses $[-3,3]$, and column (3) uses $[-4,4]$. The regression sample is drawn from the 2015, 2017, 2019, and 2021 waves of the Political Culture Survey conducted by DANE. All specifications include cohort-by-region, high school-by-region, and survey year-by-region fixed effects, as well as a dummy variable for ethnic-minority status. Robust standard errors, clustered at the region level, are reported in parentheses, and wild-bootstrap confidence intervals are shown in square brackets. *, **, and *** indicate statistical significance at the 10\%, 5\%, and 1\% levels, respectively.} }
\end{tabular}
}
\end{center}
\end{table}


\begin{figure}[h!]
             \caption{Effect of Exposure to Courses on the 1991 Constitution on Catholic Identification: Robustness to Alternative Age Windows}
        \label{fig_efftrcatholi_agewin}
\includegraphics[width=14cm,height=9cm]{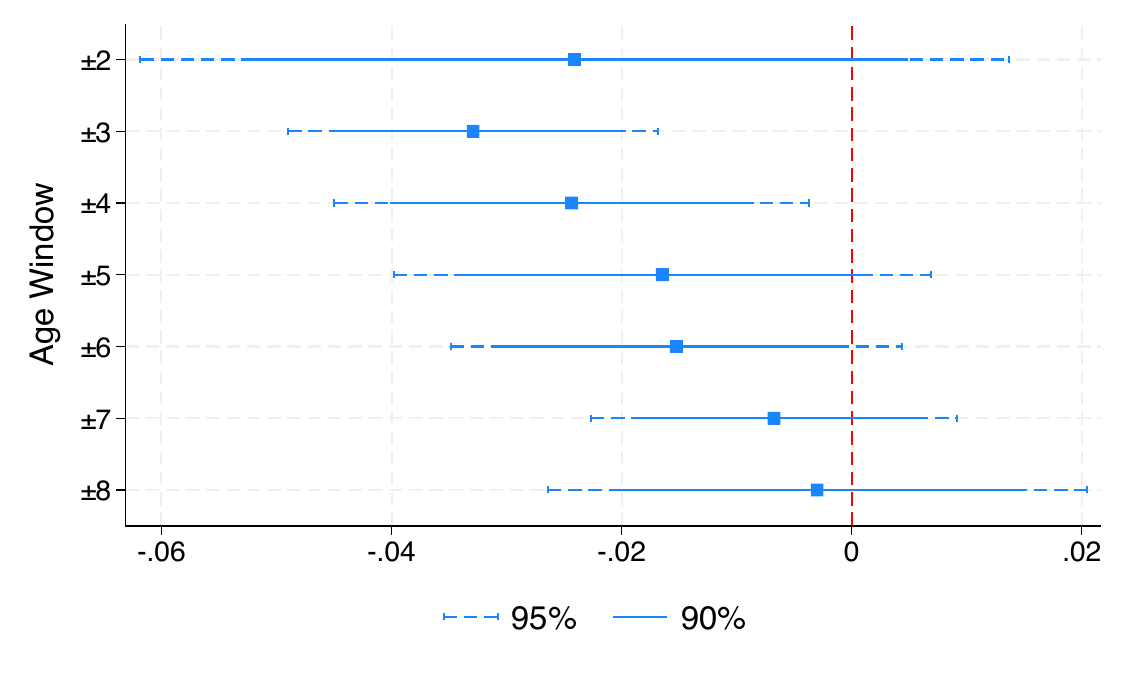}
     \begin{minipage}{16cm} \scriptsize \textbf{Note:}  This figure shows the effect of exposure to mandatory high school courses on the 1991 Constitution on Catholic identification, examining robustness to alternative age windows. The estimates correspond to Equation (\ref{baselineDiD}). The regression sample is drawn from the 2015, 2017, 2019, and 2021 waves of the Political Culture Survey conducted by DANE. All specifications include cohort-by-region, high school-by-region, and survey-year-by-region fixed effects, as well as a dummy variable for ethnic-minority status.
\end{minipage}
\end{figure}

\begin{figure}[h!]
             \caption{Effect of Exposure to Courses on the 1991 Constitution on Identification with non-Catholic Christians, non-Christians and Non-religious or Atheist: Robustness to Alternative Age Windows}
        \label{fig_efftrotherall_agewin}
\begin{center}
\vspace{-0.3cm}
\begin{subfigure}{0.5\textwidth}
\caption{Non-Catholic Christian}
\includegraphics[width=8cm,height=6cm]{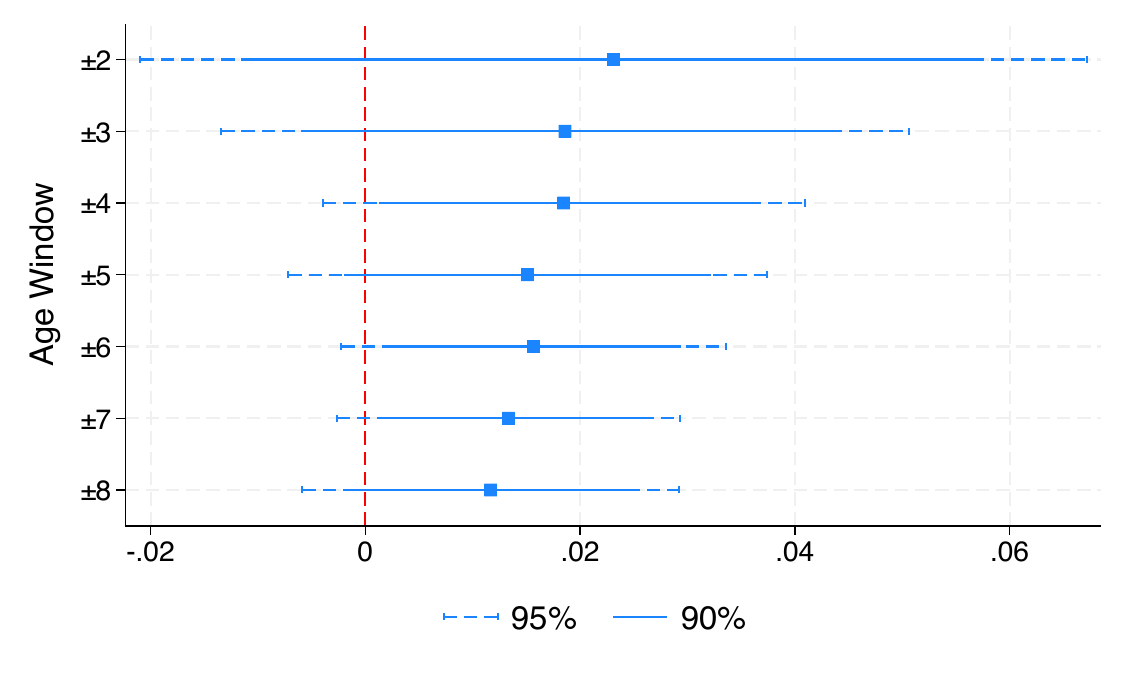}
\label{fig_efftrotherall_agewin_a}
\end{subfigure}\hspace*{\fill}
\begin{subfigure}{0.5\textwidth}
\caption{Non-Christian}
\includegraphics[width=8cm,height=6cm]{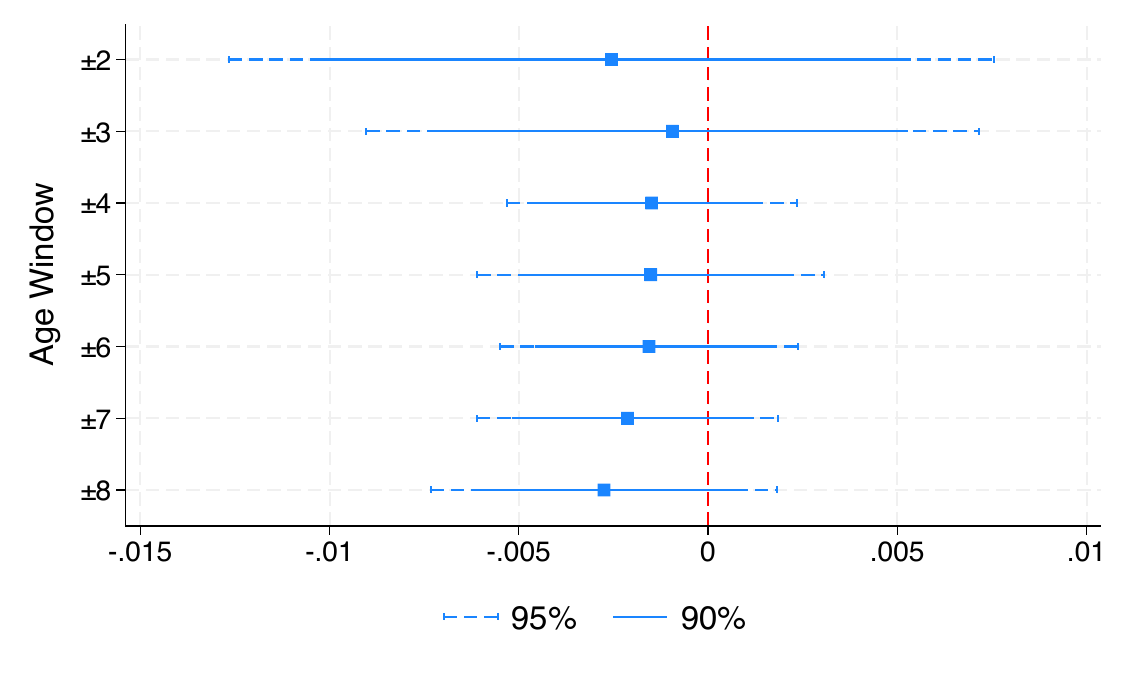}
\label{fig_efftrotherall_agewin_b}
\end{subfigure}
\begin{subfigure}{0.5\textwidth}
\caption{No religion or Atheist}
\includegraphics[width=8cm,height=6cm]{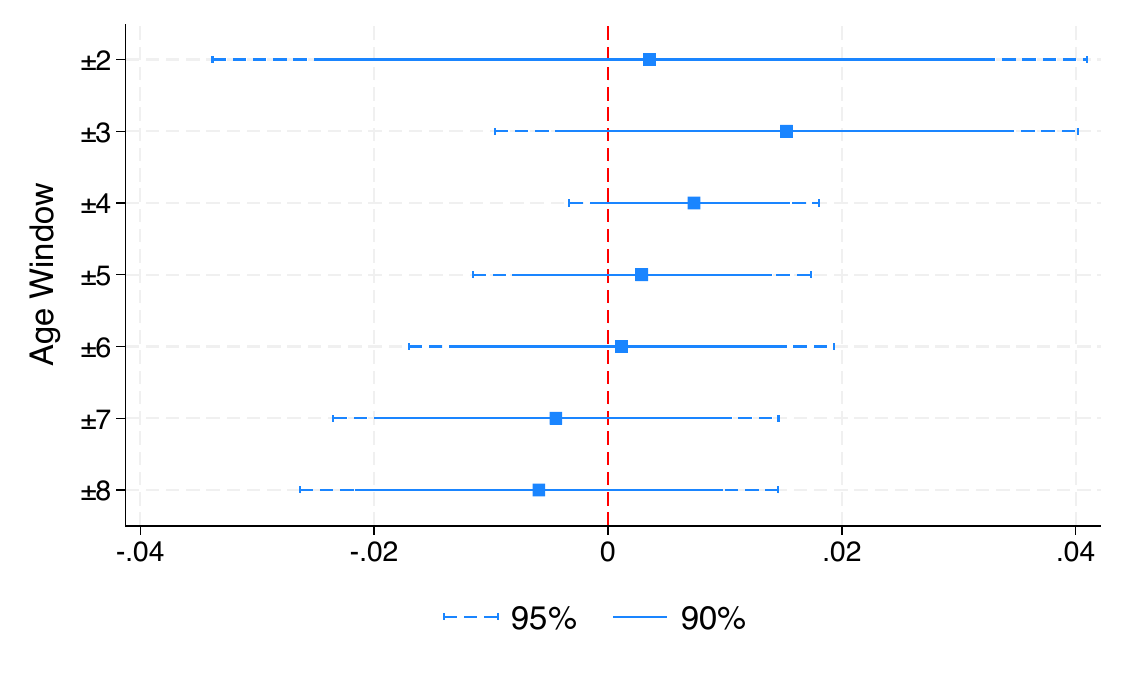}
\label{fig_efftrotherall_agewin_c}
\end{subfigure}
     \begin{minipage}{16cm} \scriptsize \textbf{Note:}  This figure shows the effect of exposure to mandatory high school courses on the 1991 Constitution on (a) identification with non-Catholic Christian denominations, (b) identification with non-Christian religions, and (c) identification with no religion, agnosticism, or atheism, examining robustness to alternative age windows. All estimates correspond to Equation~(\ref{baselineDiD}). The regression sample is drawn from the 2015, 2017, 2019, and 2021 waves of the Political Culture Survey conducted by DANE. All specifications include cohort-by-region, high school-by-region, and survey-year-by-region fixed effects, as well as a dummy variable for ethnic-minority status.
\end{minipage}
\end{center}
\end{figure}


\begin{landscape}
\begin{table}[H]
\vspace{1cm}
\begin{center}
{
\renewcommand{\arraystretch}{0.6}
\setlength{\tabcolsep}{5pt}
\caption {Effect of Exposure to Courses on the 1991 Constitution on Religious Identification: Robustness to Alternative Age Windows}  \label{tab_efftrotherall_agewin}
\vspace{-0.3cm}
\scriptsize
 \begin{tabular}{lccccccc}
\hline\hline \addlinespace[0.15cm]
    & (1)& (2)& (3)  & (4)& (5)& (6) & (7)\\\addlinespace[0.15cm]\cmidrule[0.2pt](l){2-8} \addlinespace[0.10cm]   
      & \multicolumn{1}{c}{[-2,2]} & \multicolumn{1}{c}{[-3,3]}& \multicolumn{1}{c}{[-4,4]}& \multicolumn{1}{c}{[-5,5]} & \multicolumn{1}{c}{[-6,6]}& \multicolumn{1}{c}{[-7,7]}& \multicolumn{1}{c}{[-8,8]}\\\addlinespace[0.15cm]\hline\addlinespace[0.15cm]
    \multicolumn{1}{l}{\emph{\underline{Panel A}:  }}   & \multicolumn{7}{c}{Dependent Variable: Catholic }\\\addlinespace[0.15cm]\cmidrule[0.2pt](l){2-8} \addlinespace[0.10cm]   

(Age $\leq$ 16 in 1994) $\times$ (Highschool) & -0.024 & -0.033 & -0.024 & -0.016 & -0.015 & -0.007 & -0.003 \\
  & (0.014) & (0.006)*** & (0.007)** & (0.008) & (0.007)* & (0.006) & (0.008) \\
  & [-0.069, 0.011] & [-0.046, -0.021]*** & [-0.054, -0.005]*** & [-0.057, 0.005]* & [-0.033, 0.008] & [-0.017, 0.012] & [-0.025, 0.025] \\
R-sq  & 0.019 & 0.021 & 0.021 & 0.019 & 0.019 & 0.020 & 0.019 \\
Observations  & 14,090 & 19,875 & 25,364 & 31,034 & 36,495 & 42,200 & 47,827 \\
\addlinespace[0.15cm]\hline\addlinespace[0.15cm]

    \multicolumn{1}{l}{\emph{\underline{Panel B}:  }}   & \multicolumn{7}{c}{Dependent Variable: Non-Catholic Christian}\\\addlinespace[0.15cm]\cmidrule[0.2pt](l){2-8} \addlinespace[0.10cm]   
  
(Age $\leq$ 16 in 1994) $\times$ (Highschool) & 0.023 & 0.019 & 0.018 & 0.015 & 0.016 & 0.013 & 0.012 \\
  & (0.016) & (0.012) & (0.008)* & (0.008) & (0.006)* & (0.006)* & (0.006) \\
  & [-0.029, 0.064] & [-0.014, 0.051] & [-0.008, 0.039] & [-0.001, 0.042] & [-0.034, 0.037]* & [-0.057, 0.023]* & [-0.051, 0.033] \\
R-sq  & 0.016 & 0.017 & 0.016 & 0.015 & 0.015 & 0.015 & 0.014 \\
Observations  & 14,090 & 19,875 & 25,364 & 31,034 & 36,495 & 42,200 & 47,827 \\
\addlinespace[0.15cm]\hline\addlinespace[0.15cm]

    \multicolumn{1}{l}{\emph{\underline{Panel C}:  }}   & \multicolumn{7}{c}{Dependent Variable: Non-Christian}\\\addlinespace[0.15cm]\cmidrule[0.2pt](l){2-8} \addlinespace[0.10cm]   

(Age $\leq$ 16 in 1994) $\times$ (Highschool) & -0.003 & -0.001 & -0.001 & -0.002 & -0.002 & -0.002 & -0.003 \\
  & (0.004) & (0.003) & (0.001) & (0.002) & (0.001) & (0.001) & (0.002) \\
  & [-0.012, 0.006] & [-0.006, 0.007] & [-0.006, 0.012] & [-0.006, 0.010] & [-0.005, 0.006] & [-0.007, 0.006] & [-0.006, 0.003] \\
R-sq  & 0.005 & 0.005 & 0.005 & 0.005 & 0.004 & 0.004 & 0.004 \\
Observations  & 14,090 & 19,875 & 25,364 & 31,034 & 36,495 & 42,200 & 47,827 \\
\addlinespace[0.15cm]\hline\addlinespace[0.15cm]

    \multicolumn{1}{l}{\emph{\underline{Panel D}:  }}   & \multicolumn{7}{c}{Dependent Variable: No religious or Atheist}\\\addlinespace[0.15cm]\cmidrule[0.2pt](l){2-8} \addlinespace[0.10cm]   

(Age $\leq$ 16 in 1994) $\times$ (Highschool) & 0.004 & 0.015 & 0.007 & 0.003 & 0.001 & -0.004 & -0.006 \\
  & (0.013) & (0.009) & (0.004) & (0.005) & (0.007) & (0.007) & (0.007) \\
  & [-0.028, 0.032] & [-0.001, 0.045]* & [-0.003, 0.040]* & [-0.011, 0.043] & [-0.021, 0.052] & [-0.027, 0.038] & [-0.030, 0.024] \\
R-sq  & 0.014 & 0.014 & 0.014 & 0.015 & 0.014 & 0.016 & 0.017 \\
Observations  & 14,090 & 19,875 & 25,364 & 31,034 & 36,495 & 42,200 & 47,827 \\

\addlinespace[0.15cm]\hline\hline\addlinespace[0.15cm]
\multicolumn{8}{p{21cm}}{\scriptsize{\textbf{Note:} This table reports the estimated effect of exposure to mandatory high school courses on the 1991 Constitution on four outcomes: Panel A presents results for Catholic identification, Panel B for non-Catholic Christian identification, Panel C for identification with non-Christian religions, and Panel D for identification with no religion, agnosticism, or atheism. Columns (1)–(7) report estimates of Equation~(\ref{baselineDiD}) using progressively wider symmetric age windows: column (1) uses $[-2,2]$, column (2) uses $[-3,3]$, column (3) uses $[-4,4]$, column (4) uses $[-5,5]$, column (5) uses $[-6,6]$, column (6) uses $[-7,7]$, and column (7) uses $[-8,8]$. The regression samples are drawn from the 2015, 2017, 2019, and 2021 waves of the Political Culture Survey conducted by DANE. All specifications include cohort-by-region, high school-by-region, and survey-year-by-region fixed effects, as well as a dummy variable for ethnic-minority status. Robust standard errors, clustered at the region level, are reported in parentheses, and wild-bootstrap confidence intervals are shown in square brackets. *, **, and *** indicate statistical significance at the 10\%, 5\%, and 1\% levels, respectively.} }
\end{tabular}
}
\end{center}
\end{table}
\end{landscape}

\begin{table}[H]
\vspace{3cm}
\begin{center}
{
\renewcommand{\arraystretch}{0.8}
\setlength{\tabcolsep}{3pt}
\caption {Effect of Exposure to 1991 Constitution Courses on Religious Identification: Excluding Bogota}  \label{tab_effectallregionnobogota}
\vspace{-0.3cm}
\scriptsize
 \begin{tabular}{lccccc}
\hline\hline \addlinespace[0.15cm]
    & (1)& (2)& (3)  & (4)& (5)\\\addlinespace[0.12cm]\cmidrule[0.2pt](l){2-6}\addlinespace[0.12cm]
& \multicolumn{5}{c}{Dependent Variable: Identification with the following groups}\\\addlinespace[0.15cm]\cmidrule[0.2pt](l){2-6} \addlinespace[0.10cm]   

&\multicolumn{1}{c}{Catholic}&\multicolumn{1}{c}{Non-Catholic}&\multicolumn{1}{c}{Non-Christian}&\multicolumn{1}{c}{No Religion}&\multicolumn{1}{c}{Religious}\\

&\multicolumn{1}{c}{}&\multicolumn{1}{c}{Christian}&\multicolumn{1}{c}{}&\multicolumn{1}{c}{or Atheist}&\multicolumn{1}{c}{attendance}\\

\addlinespace[0.15cm]\hline\addlinespace[0.15cm]

\multicolumn{1}{l}{\emph{\underline{Panel A}:  }}      & \multicolumn{5}{c}{Baseline (all regions)}\\\addlinespace[0.1cm]\cmidrule[0.2pt](l){2-6}\addlinespace[0.1cm]  
(Age $\leq$ 16 in 1994) $\times$ (High school) & -0.033 & 0.020 & -0.001 & 0.014 & -0.025 \\
  & (0.006)** & (0.013) & (0.003) & (0.010) & (0.038) \\
  & [-0.046, -0.020]*** & [-0.013, 0.053] & [-0.006, 0.007] & [-0.004, 0.047] & [-0.131, 0.044] \\
R-sq  & 0.024 & 0.018 & 0.004 & 0.015 & 0.046 \\
Observations  & 16,999 & 16,999 & 16,999 & 16,999 & 6,613 \\

\addlinespace[0.18cm]\hline\addlinespace[0.12cm]

\multicolumn{1}{l}{\emph{\underline{Panel B}:  }}      & \multicolumn{5}{c}{Central and East Regions}\\\addlinespace[0.1cm]\cmidrule[0.2pt](l){2-6}\addlinespace[0.1cm]  
(Age $\leq$ 16 in 1994) $\times$ (High school) & -0.033 & 0.007 & -0.004 & 0.030 & -0.079 \\
  & (0.010) & (0.021) & (0.002) & (0.013) & (0.045) \\
  & [-0.043, -0.021]*** & [-0.017, 0.028] & [-0.006, -0.002]*** & [0.015, 0.044]*** & [-0.128, -0.036]*** \\
R-sq  & 0.015 & 0.011 & 0.005 & 0.015 & 0.034 \\
Observations  & 7,666 & 7,666 & 7,666 & 7,666 & 3,127 \\

\addlinespace[0.18cm]\hline\addlinespace[0.12cm]

\multicolumn{1}{l}{\emph{\underline{Panel C}:  }}      & \multicolumn{5}{c}{Ethnic Minority}\\\addlinespace[0.1cm]\cmidrule[0.2pt](l){2-6}\addlinespace[0.1cm]  
(Age $\leq$ 16 in 1994) $\times$ (High school) & -0.038 & 0.008 & 0.014 & 0.015 & 0.181 \\
  & (0.027) & (0.028) & (0.008) & (0.016) & (0.009)*** \\
  & [-0.119, 0.081] & [-0.024, 0.060] & [-0.069, 0.028] & [-0.077, 0.130] & [0.147, 0.313]*** \\
R-sq  & 0.028 & 0.023 & 0.014 & 0.031 & 0.097 \\
Observations  & 3,143 & 3,143 & 3,143 & 3,143 & 976 \\

\addlinespace[0.18cm]\hline\addlinespace[0.12cm]

\multicolumn{1}{l}{\emph{\underline{Panel D}:  }}      & \multicolumn{5}{c}{No Ethnic Minority}\\\addlinespace[0.1cm]\cmidrule[0.2pt](l){2-6}\addlinespace[0.1cm]  
(Age $\leq$ 16 in 1994) $\times$ (High school) & -0.030 & 0.021 & -0.005 & 0.013 & -0.062 \\
  & (0.009)** & (0.015) & (0.003) & (0.008) & (0.029) \\
  & [-0.045, -0.007]*** & [-0.021, 0.050] & [-0.017, 0.000]*** & [-0.007, 0.031] & [-0.143, -0.017]*** \\
R-sq  & 0.020 & 0.016 & 0.005 & 0.013 & 0.049 \\
Observations  & 13,856 & 13,856 & 13,856 & 13,856 & 5,637 \\

\addlinespace[0.15cm]\hline\hline\addlinespace[0.15cm]
\multicolumn{6}{p{16.5cm}}{\scriptsize{\textbf{Note:} This table reports estimates of the effect of exposure to mandatory high school courses on the 1991 Constitution on religious identification, distinguishing between Catholic, non-Catholic Christian, non-Christian, and secular affiliations. All estimations exclude Bogota from the sample. Panel B focuses on individuals residing in the Central and Eastern regions, while Panels C and D restrict the sample to individuals who self-identify as ethnic minorities. All columns report estimates of Equation (\ref{baselineDiD}) using an age window of $[-3, 3]$. The regression sample is drawn from the 2015, 2017, 2019, and 2021 waves of the Political Culture Survey conducted by DANE. All specifications include cohort, high school, and survey fixed effects, as well as dummy variables for female and ethnic-minority status. Robust standard errors, clustered at the region level, are reported in parentheses, and wild-bootstrap confidence intervals are shown in square brackets.  *, **, and *** indicate statistical significance at the 10\%, 5\%, and 1\% levels, respectively.} }
\end{tabular}
}
\end{center}
\end{table}

\end{document}